\documentclass{article}
\usepackage[utf8]{inputenc}
\usepackage{amsmath,amssymb,amsfonts}
\usepackage{algorithmic}
\usepackage[linesnumbered,ruled,vlined]{algorithm2e}

\SetCommentSty{mycommfont}
\usepackage{graphicx}
\usepackage{textcomp}
\usepackage{pgfplots}
\pgfplotsset{compat=1.14}
\usepackage{xcolor}
\usepackage{amsthm}
\usepackage{multirow}
\usepackage{breqn}
\usepackage{subcaption}
\usepackage[numbers]{natbib}
\usepackage{graphicx}
\usepackage{url}
\usepackage[affil-it]{authblk}

\providecommand{\keywords}[1]
{
  \small	
  \textbf{\textit{Keywords---}} #1
}

\begin{document}


\title{HiCOPS: High Performance Computing Framework for Tera-Scale Database Search of Mass Spectrometry based Omics Data}

\author{Muhammad Haseeb$^*$, Fahad Saeed\thanks{School of Computing, and Information Sciences, Florida International University (FIU), Miami, FL USA; Corresponding Author. Email: fsaeed@fiu.edu}}
\date{}
\maketitle

%
\vspace{-1cm}
\begin{abstract}
Database-search algorithms, that deduce peptides from Mass Spectrometry (MS) data, have tried to improve the \emph{computational efficiency} to accomplish larger, and more complex systems biology studies. Existing serial, and high-performance computing (HPC) search engines, otherwise highly successful, are known to exhibit poor-scalability with increasing size of theoretical search-space needed for increased complexity of modern non-model, multi-species MS-based omics analysis. Consequently, the bottleneck for computational techniques is the communication costs of moving the data between hierarchy of memory, or processing units, and not the arithmetic operations.  This post-Moore change in architecture, and demands of modern systems biology experiments have dampened the overall effectiveness of the existing HPC workflows.
We present a novel efficient parallel computational method, and its implementation on memory-distributed architectures for peptide identification tool called HiCOPS, that enables more than 100-fold improvement in speed over most existing HPC proteome database search tools. HiCOPS empowers the supercomputing database search concept for comprehensive identification of peptides, and all their modified forms within a reasonable time-frame. We demonstrate this by searching Gigabytes of experimental MS data against Terabytes of databases where HiCOPS completes peptide identification in few minutes using 72 parallel nodes (1728 cores) compared to several weeks required by existing state-of-the-art tools using 1 node (24 cores); 100 minutes vs 5 weeks; 500$\times$ speedup. Finally, we formulate a theoretical framework for our \emph{overhead-avoiding} strategy, and report superior performance evaluation results for key metrics including execution time, CPU utilization, speedups, and I/O efficiency. We also demonstrate superior performance as compared to all existing HPC strategies. 
\end{abstract}


%
%
\section*{Main}

Faster, and more efficient peptide identification algorithms \cite{nesvizhskii2010survey} \cite{kong2017msfragger} \cite{mcilwain2014crux} have been the cornerstone of computational research in shotgun MS based proteomics for more than 30 years \cite{yuan2012pparse,deng2019pclean,degroeve2013ms2pip,zhou2017pdeep, mcilwain2014crux, zhang2012peaks,devabhaktuni2019taggraph,chi2018open, kong2017msfragger, bern2007lookup,  eng1994approach, craig2003method, diament2011faster, eng2008fast, park2008rapid, geer2004open}. Millions of raw, noisy spectra can be produced, in a span of few hours, using modern mass spectrometry technologies producing several gigabytes of data~\cite{hebert2014one} (Supplementary Fig.~1). Database peptide search is the most commonly employed computational approach to identify the peptides from the experimental spectra \cite{nesvizhskii2006dynamic}, \cite{chi2018open}, \cite{kong2017msfragger}, \cite{eng2011face}. In this approach, the experimental spectra are searched against a database of model-spectra (or theoretical-spectra) with the goal to find the best possible matches \cite{nesvizhskii2010survey}. The model-spectra database is simulated through in-silico techniques using a proteome sequence database (Supplementary Fig.~2). The model-spectra database can grow exponentially in space (several giga to terabytes) as the post-translational modifications (PTMs) are incorporated in simulation \cite{kong2017msfragger}, \cite{haseeb2019efficient}. Therefore, the cost of moving, and managing this data to match with the spectra now \emph{exceeds} the costs of doing the arithmetic operations in these search engines leading to non-scalable workflows with increasingly larger, and complex data sets \cite{saeed2020communication}. 

As demonstrated by other big data fields \cite{marx2013biology}, such limitations can be reduced by developing parallel algorithms that combine the computational power of thousands of processing elements across distributed-memory clusters, and supercomputers. We, and others have developed high-performance computing (HPC) techniques for processing of MS data including for multicore \cite{mcilwain2014crux}, \cite{kong2017msfragger}, \cite{chi2018open}, \cite{devabhaktuni2019taggraph}, and distributed-memory architectures~\cite{duncan2005parallel}, \cite{bjornson2007x} \cite{pratt2011mr}, \cite{li2019mctandem}, \cite{li2019sw}, \cite{prakash2019bolt}. Similar to serial algorithms, the objective of these HPC methods has been to speed up the arithmetic scoring part of the search engines, by spawning multiple (managed) instances of the original code, replicating the theoretical database, and splitting the experimental data. However, computationally optimal HPC algorithms that minimize both the computational and communications costs for these tasks are still needed. Urgent need for developing methods that exhibit optimal performance is illustrated in our theoretical framework \cite{saeed2020communication}, and can \emph{potentially} lead to large-scale systems biology studies especially for meta-proteomics, proteogenomic, and MS based microbiome or non-model organisms’ studies having direct impact on personalized nutrition, microbiome research, and cancer therapeutics.

In our quest to develop faster strategies applicable to MS based omics data analysis, we designed a novel HPC framework that provides orders-of-magnitude faster processing over \emph{both} serial, and parallel tools. We implemented this framework in a new HPC tool, capable of scaling on large (distributed) symmetric multiprocessor (SMP) supercomputers, called \emph{HiCOPS}. HiCOPS makes searches possible (in few minutes) even for tera-byte level theoretical database(s); something not feasible (several \emph{weeks} of computations) with existing state-of-the-art methods. We demonstrate HiCOPS's utility in both closed- and open-searches across different search-parameters, and experimental conditions. Further, our experimental results depict more than 100$\times$ speedup for HiCOPS compared to several existing shared and distributed memory database peptide search tools. HiCOPS is overhead-avoiding strategy that splits the database (algorithmic workload) among the parallel processes in a load balanced fashion, executes the partial database peptide search, and merges the results in communication optimal way thereby, alleviating the resource upper bounds that exist in the current generation of database peptide search tools. 

We demonstrate HiCOPS results on several data- and compute-intensive experimental conditions including using 4TB of theoretical database against which millions of spectra were matched. We show that HiCOPS even when using similar scoring functions outperforms both parallel, and serial methods. Although not a fair comparison, one of our experiments  of searching 41GB experimental spectra against a database size of 1.8TB ran in only 103.5 minutes using 72 parallel nodes compared to MSFragger which took about 35.5 days to complete the same experiment on 1 node (494$\times$ slower). HiCOPS completed an open-search (dataset size:  8K spectra, database size:  93.5M spectra) in 144 seconds as compared to the X!!Tandem (33 minutes) and SW-Tandem (4.2 hours), all using 64 parallel nodes; demonstrating that HiCOPS was also out-performing existing parallel tools. We designed 12 different experiment sets and demonstrate the performance our parallel computing framework. Our extensively evaluated HiCOPS parallel performance using metrics such as parallel efficiency: 70-80\%, load imbalance cost: $\leq$10\%, CPU utilization: improved with parallel nodes, communication costs: $\leq$10\%, I/O costs: $\leq$5\% and task scheduling related costs: $\leq$ 5\%; demonstrate superior performance as compared to any of the existing serial or parallel solution. HiCOPS is not limited to data from a particular MS instrument, allows searches on multiple model species databases, and can be incorporated into existing data analysis pipelines. HiCOPS is the first software pipeline capable of efficiently scaling to the terabyte-scale workflows using large number of parallel nodes in database peptide search domain.

%
%
\section*{Results}
\begin{figure}[htbp]
\centerline{\includegraphics[width=0.95\linewidth]{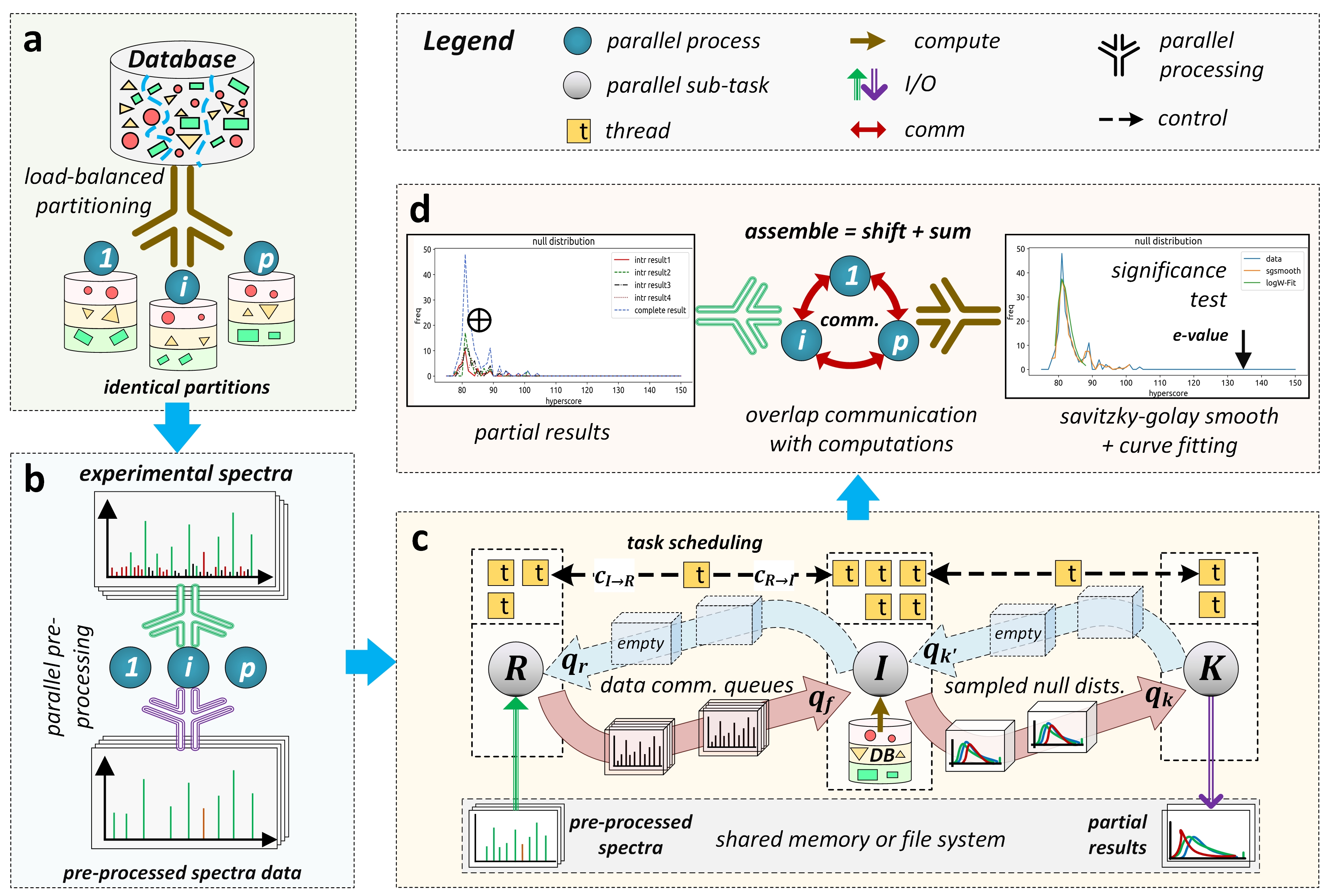}}
\caption{ \textbf{(a) Superstep 1:} The massive model-spectra database (shown as shapes) is partitioned among parallel MPI processes in load balanced manner and then locally indexed. \textbf{(b) Superstep 2:} The experimental MS/MS spectra data are split, indexed, tagged, pre-processed and written back to the file system in parallel. \textbf{(c) Superstep 3:} The partial database peptide search pipeline executed by all parallel processes is shown. On each process, three parallel sub-tasks $R$, $I$ and $K$ work in producer-consumer pipeline to load the pre-processed data, execute the partial database search producing partial results, and write the (sampled) results to the shared memory respectively. The available threads are managed between parallel sub-tasks through a task scheduling algorithm. The sub-tasks communicate via buffer queues to avoid fragmentation. \textbf{(d) Superstep 4:} The partial results are assembled into complete results to compute statistical scores which are communicated to their origin processes.}
\label{fig:grabs}
\end{figure}

HiCOPS constructs the parallel database peptide search algorithmic workflow (task-graph) using \emph{four} Single Program Multiple Data (SPMD) Bulk Synchronous Parallel (BSP) \cite{valiant1990bridging} supersteps; where a set of processes ($p_{i}\  \epsilon\  P$) execute ($\phi$) supersteps in asynchronous parallel fashion and synchronize between them. As shown in Fig ~\ref{fig:grabs}, HiCOPS allows searching of \emph{partial} theoretical database, in parallel; something that has not been accomplished in the context of peptide database-search tools. These partial search-results are then merged using a communication-optimal technique. 

\begin{figure}[htpb]
    \centering
    \includegraphics[width=0.95\textwidth]{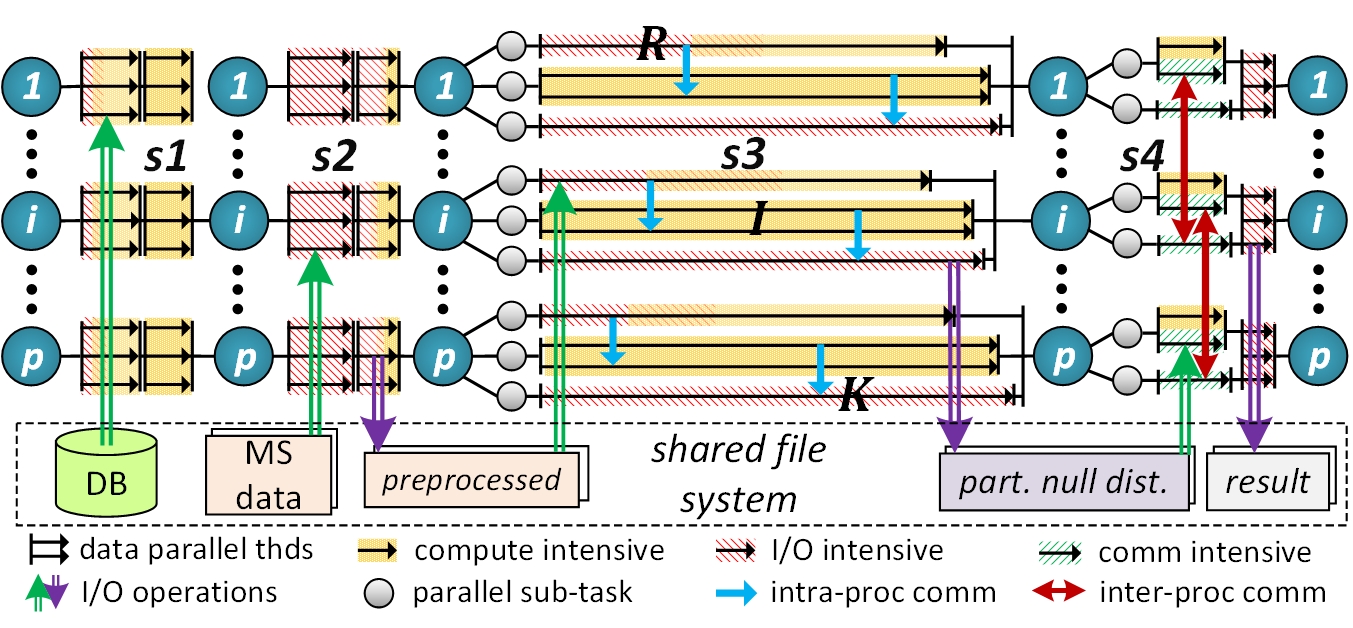}
    \caption{\textbf{Workload Profile:} Supersteps 1 and 2 are designed as data parallel. Supersteps 3 and 4 are designed as hybrid task and data parallel. The workload executed by the four respective supersteps are compute intensive, I/O intensive, mixed (compute and I/O), and mixed (compute and comm.). In the last two supersteps, the compute workload may supersede the communication and/or I/O, given that the associated overhead costs are overlapped or minimized.}
    \label{fig:main}
\end{figure}

In the first superstep, the massive model-spectra database is partitioned across parallel processes in a load balanced fashion. In the second superstep, the experimental data are divided into batches and pre-processed if required. In the third superstep, the parallel processes execute a partial database peptide search on the pre-processed experimental data batches, producing intermediate results. In the final superstep, these intermediate results are de-serialized and assembled into complete (global) results. The statistical significance scores are computed (Online Methods, Fig.~\ref{fig:grabs}) using global results. Fig.~\ref{fig:main}  gives an overview of the parallelization scheme, task-graph, and workload profile for each of the HiCOPS' supersteps (Online Methods).

The total wall time ($T_{H}$) for executing the four supersteps is the sum of superstep execution times, given as:
$$
    T_{H} = T_{1} + T_{2} + T_{3} + T_{4}
$$

Where the execution time for a superstep ($j$) is the maximum time required by any parallel task ($p_i \ \epsilon \ P$) to complete that superstep, given as:
$$
    T_{j} = max(T_{j,p_{1}}, T_{j,p_{2}}, ..., T_{j,p_{P}})
$$

Or simply:
$$
    T_{j} = max_{p_{i}}(T_{j,p_{i}})
$$

Combining the above three equations, the total HiCOPS runtime is given as:
\begin{dmath} \label{eq:tot}
    T_{H} = \sum_{j=1}^{4}{max_{p_{i}}(T_{j,p_{i}})}
\end{dmath}

%
%
\subsection*{Experimental Setup}
We used the following datasets from Pride Archive for experimentation and evaluation purposes. 
\begin{itemize}
    \item $E_{1}$: PXD009072 (0.305 million spectra)
    \item $E_{2}$: PXD020590 (1.6 million spectra)
    \item $E_{3}$: PXD015890 (3.8 million spectra)
    \item $E_{4}$: PXD007871, 009072, 010023, 012463, 013074, 013332, 014802, and 015391 combined (1.515 million spectra)
    \item $E_{5}$: All above datasets combined (6.92 million spectra)
\end{itemize} 

The search experiments were conducted against the following protein sequence databases. The databases were digested in-silico using Trypsin as enzyme with 2 allowed missed cleavages, peptide lengths between 6 and 46 and peptide masses between 500 and 5000Da. The number and type of PTMs added to the database, and the peptide precursor mass tolerance ($\delta M$) were varied across experiments however, the fragment mass tolerance ($dF$) was set to $\pm$0.005Da in all experiments.
\begin{itemize}
    \item $D_1$: UniProt Homo sapiens (UP005640)
    \item $D_2$: UniProt SwissProt (reviewed, \emph{multi-species})
\end{itemize}

Furthermore, we designed 12 different experiments ($e_{n}$) using combinations of the above mentioned databases, datasets and experimental parameters for our extensive performance evaluation. These experiments exhibit varying experimental workloads to cover a wide-range of real-world scenarios. We represent each of these experiment sets using a tuple: $e_{n} = (q,D,\delta M)$ where $q$ is dataset size in 1 million spectra, $D$ is model-spectra database size in 100 million spectra and $\delta M$ peptide precursor mass setting in $\pm$100Da to represent the problem size. The designed experiment sets (of varying workloads) are listed as: $e_1$=(0.3, 0.84, 0.1), $e_2$=(0.3, 0.84, 2), $e_3$=(3.89, 0.07, 5), $e_4$=(1.51, 2.13, 5), $e_5$=(6.1, 0.93, 5), $e_6$=(3.89, 7.66, 5), $e_7$=(1.51, 19.54, 5), $e_8=$(1.6, 38.89, 5), $e_9$=(3.89, 15.85, 5), $e_{10}$=(3.89, 1.08, 5), $e_{11}$=(1.58, 2.13, 1), and $e_{12}$=(0.305, 0.847, 5).

\textbf{Runtime Environment:} All distributed memory tools were run on the Extreme Science and Engineering Discovery Environment (XSEDE) \cite{towns2014xsede} Comet cluster at the San Diego Supercomputer Center (SDSC). All Comet compute nodes are equipped with 2 sockets $\times$ 12 cores of Intel Xeon E5-2680v3 processor, 2 NUMA nodes $\times$ 64GB DRAM, 56 Gbps FDR InfiniBand interconnect and Lustre shared file system. The maximum number of nodes allowed per job is 72 and maximum allowed job time is 48 hours. The shared memory tools, on the other hand, were run on a local server system equipped with an Intel Xeon Gold 6152 processor (22 physical cores, 44 hardware threads), 128GB DRAM and a local 6TB SSD storage as most experiments using the shared memory tools required $>$48 hours (job time limit on XSEDE Comet). 

%
%
\subsection*{Correctness of the parallel design}
We evaluated the correctness of our parallel design by searching all five datasets $E_{i}$ against both protein sequence databases $D_{i}$ under various settings, and combinations of PTMs. The correctness was evaluated in terms of of consistency in the number of database hits, the identified peptide to spectrum matches (PSM), and the hyper-scores and e-values assigned to those sequences (within 3 decimal points) for each experimental spectrum searched. The experiments were performed using combinations of experimental settings where we observed more than 99.5\% consistent results regardless of the number of parallel nodes. The negative error in expected values results observed in erroneous identifications was caused by the sampling, and floating-point precision losses (Online Methods, Fig.~\ref{fig:sampling}, Fig.~\ref{fig:grabs}d). A snippet of the 251,501 peptide to spectrum match (PSM) results obtained by searching the dataset: $E_{1}$ against the database: $D_{1}$ with no post-translational modifications added at precursor mass tolerance: $\delta M$=$\pm$500Da is shown in Supplementary Table~1.

%
%
\subsection*{Comparative analysis reveals orders of magnitude speedups}
We compared the HiCOPS speed against many existing shared and distributed memory parallel database peptide search algorithms including MSFragger v3.0 \cite{kong2017msfragger}, X! Tandem v17.2.1 \cite{craig2004tandem}, Tide/Crux v3.2 \cite{mcilwain2014crux}, X!! Tandem v10.12.1 \cite{bjornson2007x}, and SW-Tandem \cite{li2019sw}. In the first experiment set, a subset of 8,000 spectra (file: 7Sep18\_Olson\_WT24) from dataset: $E_{3}$ was searched against the database: $D_{2}$. Fixed Cysteine Carbamidomethylation, and variable Methionine oxidation, and Tyrosine Biotin-tyramide were added yielding model-spectra database of 93.5 million ($\sim$90GB). In the second experiment set, the entire dataset: $E_{3}$ was searched against the same database $D_{2}$. The peptide precursor mass tolerance was in both sets was first set to: $\delta M$=$\pm$10Da and then $\pm$500Da ($\pm$100Da for Tide/Crux). The obtained wall time results (Supplementary Tables~2, 3, 4, 5) show that HiCOPS outperforms both the shared and distributed memory tools (in speed) by $>$100$\times$ when the experiment size is large. For instance, for the second experiment, HiCOPS outperforms both the X!!Tandem and SW-Tandem by $>800\times$ (230seconds v $>$2days) using 64 nodes. HiCOPS also depicts speedups of 67.3$\times$ and 350$\times$ versus MSFragger (1 node) for the same experiment set. Furthermore, we observed no speedups for SW-Tandem with increasing number of parallel nodes (no parallel efficiency). We repeatedly contacted the corresponding authors about the parallel efficiency issue but did not receive a response (Supplementary Text~6).

%
%
\subsection*{Application in large-scale peptide identification}
The application of HiCOPS in extremely resource intensive experimental settings was demonstrated using additional experiments where the datasets: $E_{3}$, $E_{4}$ and $E_{2}$ were searched against model-spectra databases of sizes: 766M (780GB), 1.59B (1.7TB) and 3.88B (4TB) respectively ($\delta M$=$\pm$500Da). HiCOPS completed the execution of these experiments using 64 parallel nodes (1538 cores) in 14.55 minutes, 103.5 minutes and 27.3 minutes respectively. To compare, we ran the second experiment (dataset: $E_{4}$ and database size: 1.59B (1.7TB)) on MSFragger which completed after 35.5 days making HiCOPS 494$\times$ faster. The rest of the experiments were intentionally not run using any other tools but HiCOPS to avoid feasibility issues as each tool would require several months of processing to complete each experiment, as evident from Supplementary Tables~3, 4, 5. The wall clock execution time results for this set of experiments are summarized in the Table~\ref{table:application}.

\begin{table}[htpb!]
  \centering
  \caption{Experimental wall times for large-scale peptide identification experiments using HiCOPS and MSFragger. Exp. 1: (DB: 766M, DS: 3.8M, $\delta M$=500Da), Exp. 2: (DB: 1.6B, DS: 1.5M, $\delta M$=500Da), Exp. 3: (DB: 3.88B, DS: 1.6M, $\delta M$=500Da). The experiments were not performed using other tools due to their relatively slower speeds requiring several months of processing per tool per experiment.
  }
  \label{table:application}
  \begin{tabular}{|c|c|c|c|c|c|}
    \hline 
    \multirow{2}{*}{\parbox{1.0cm}{\centering{\textbf{Exp. Num}}}} & \multirow{2}{*}{\textbf{Tool}} & \multirow{2}{*}{\textbf{Nodes}} & \multirow{2}{*}{\parbox{1.3cm}{\centering{\textbf{Dataset (GB)}}}} & \multirow{2}{*}{\parbox{1.5cm}{\centering{\textbf{Database (GB)}}}} & \multirow{2}{*}{\parbox{1.1cm}{\centering{\textbf{Time (min)}}}}\\
     & & & & & \\
     \hline
    1 & HiCOPS & 64 & 20 & 780 & 14.55\\
    \hline
    2 & HiCOPS & 64 & 15 & 1692 & 103.5\\
    \hline
    2 & MSFragger & 1 & 15 & 1692 & 51130\\
    \hline
    3 & HiCOPS & 64 & 41 & 4000 & 27.3\\
    \hline
\end{tabular}
\end{table}

%
%
\subsection*{HiCOPS exhibits efficient strong-scale speedups}
\begin{figure}[htbp]
\centerline{\includegraphics[width=0.95\linewidth]{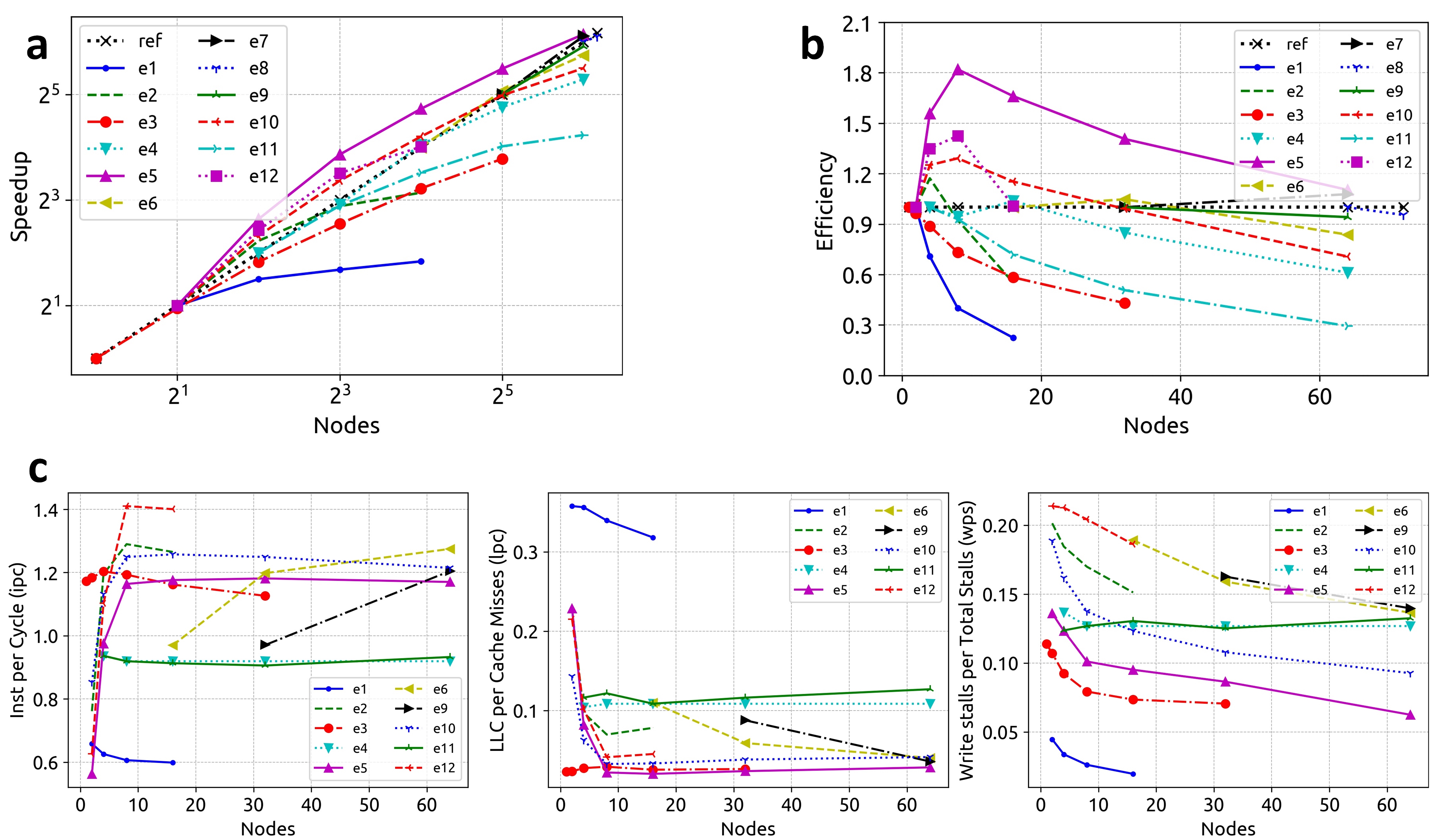}}
\caption{\textbf{(a, b)} The speedup and parallel efficiency improve as the experimental workload increases ranging between 70-80\% for most experiment sets. \textbf{(c)} The hardware utilization metrics show an improved performance per node trend for large workloads as the number of parallel nodes increase resulting in super-linear speedups (e.g. $e_5$).}
\label{fig:speedups}
\end{figure}

The speedup and strong scale efficiency for the overall and superstep-by-superstep runtime was measured for all 12 experiment sets. The results (Supplementary Fig.~4, Fig~\ref{fig:speedups}a, b) depict that the overall strong scale efficiency closely follows the superstep 3 (evident in Supplementary Fig.~4) and ranges between 70-80\% for sufficiently large experimental workload. Super-linear speedups were also observed in many experiments with higher workloads. To explain this, the following hardware counters-based metrics were also recorded for all experiment sets: instructions per cycle ($ipc$), last level cache misses per all cache level misses ($lpc$), and the cycles stalled due to writes per total stalled cycles ($wps$). The results (Fig.~\ref{fig:speedups}c) show that the CPU, cache, and memory bandwidth utilization improves as the workload per node ($wf/P$) increases reaching to an optimum point after which it saturates due to memory bandwidth contention since the database search algorithms employed (and also in general) are highly memory intensive. Beyond this saturation point, increasing the number of parallel nodes for the same experimental workload resulted in a substantial improvement (super-linear) in performance as the workload per node ($wf/P$) reduced to the normal (optimal) range. For instance, the experiment set $e_{5}$ depicts super-linear speedups (Fig.~\ref{fig:speedups}a) which can be correlated to the hardware performance surge in Fig.~\ref{fig:speedups}c.

%
%
\subsection*{Performance evaluation reveals minimal overhead costs}
\begin{figure}[htbp]
\centerline{\includegraphics[width=0.95\linewidth]{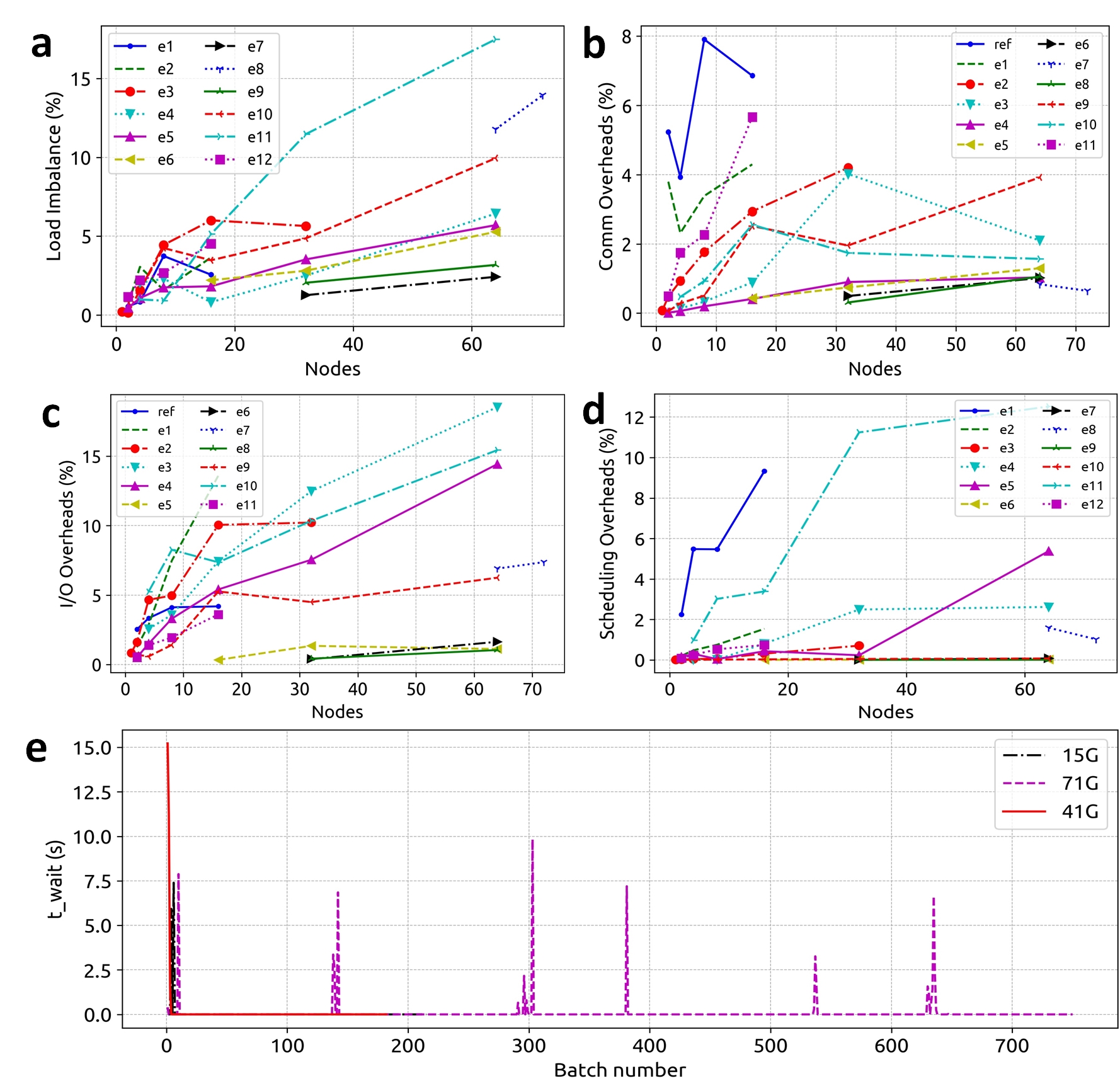}}
\caption{\textbf{(a)} The load imbalance overhead costs remain under 10\% in most experiment sets. \textbf{(b)} The communication overhead costs remain under 5\% in most experiments sets. \textbf{(c)} The I/O overheads remain under 10\% in most experiment sets but there is an upward trend as the number of parallel nodes increase. This occurs due to the saturation of the shared file system bandwidth. \textbf{(d)} The scheduling costs remain under 5\% for most experiment sets. The scheduling costs may increase if the workload per node is extremely small. \textbf{(e)} The time series shows that the task scheduling algorithm efficiently redistributes the parallel threads as soon as a surge in cost is detected.}
\label{fig:overheads}
\end{figure}

The load imbalance, communication, I/O, and task scheduling costs were measured for all experiment 12 sets. The obtained results (Fig.~\ref{fig:overheads}a, b, c) depict that the load imbalance costs remain $\leq$10\%, communication costs remain $\leq$5\%, I/O costs remain $\leq$10\% in most experiments. Note that the load imbalance is a direct measure of synchronization cost. The task-scheduling cost was measured through a time series ($t_{wait}$) (Fig.~\ref{fig:overheads}e) which monitors the time that the parallel cores had to wait for the data I/O to complete. The results (Fig.~\ref{fig:overheads}e) depict that our task-scheduling algorithm actively performs counter measures (reallocates threads) as soon as a surge in wait-time is detected keeping the cost to $\leq$ 5\% in most experiments (Fig.~\ref{fig:overheads}d). We also observed that the I/O cost is affected by a number of factors including average dataset file size, number of files in the dataset and the available file system bandwidth. The communication cost is affected by the available network bandwidth.

%
%
\section*{Discussion}
Enormous possibilities of chemical and biological modifications add knowledge discovery dimensions to mass spectrometry-based omics but are not explored in most studies, in part, due to the scalability challenges associated with comprehensive PTM searches. Current MS based computational proteomics algorithms, both serial and parallel, have focused on improving arithmetic computations by introducing indexing and approximation methods to speedup their workflows. However recent trends in the workloads stemming from systems biology (e.g. meta-proteomics, proteogenomics) experiments point towards urgent need for computational tools, capable of efficiently harnessing the compute and memory resources from supercomputers. Our highly scalable and low-overhead strategy, HiCOPS, meets this urgent need for next generation of computational solutions leading to more comprehensive peptide identification application. Further, our HPC framework can be adapted for accelerating most existing modern database peptide search algorithms.

We demonstrate using never before done experiments, peptide deduction through searching gigabytes of experimental MS/MS data against terabytes of model-spectra databases in only a few minutes compared to several days required by modern tools (100 minutes vs 5 weeks; 500$\times$ speedup using 72 parallel nodes). Our overhead-avoiding BSP-model based parallel algorithmic design allows efficient exploitation of extreme-scale resources available in modern high-performance computing architectures, and supercomputers. Extensive performance evaluation using over two dozen experiment sets with variable problem size (database and dataset sizes) and experimental settings revealed superior strong scale parallel efficiency, and minimal overhead costs for HiCOPS. HiCOPS is a novel HPC framework that gives systems biologist a tool to perform \emph{comprehensive} modification searches for meta-proteomics, proteogenomic, and proteomics studies for non-model organisms at scale. HiCOPS is under direct development and will update with improved I/O efficiency, load balancing, reduced overhead costs, and the parallel design for heterogeneous and CPU-GPU architectures in future versions. Therefore, we believe that the peptide search strategy (both open- and closed) for comprehensive PTM’s, made practical by HiCOPS, has the potential to become a valuable option for scalable analysis of shotgun Mass Spectrometry based omics.

\newpage

%
%
\section*{Online Methods}\label{sec:online_meth}
\subsection*{Notations and Symbols}
For the rest of the paper, we will denote the number of peptide sequences in the database as ($\zeta$), average number of post-translational modifications (PTMs) per peptide sequence as ($m$), the total database size as ($\zeta(2^{m})=D$), the number of parallel nodes/processes as ($P$), number of cores per parallel process as ($c_{p_{i}}$), size of experimental MS/MS dataset (i.e. number of experimental/query spectra) as ($q$), average length of query spectrum as ($\beta$), and the total dataset size as ($q\beta$). The runtime of executing the superstep ($j$) by parallel task ($p_{i}$) will be denoted as ($T_{j,p_{i}}$) and the generic overheads due to boilerplate code, OS delays, memory allocation etc. will be captured via ($\gamma_{p_{i}}$).

%
%
\subsection*{Runtime Cost Model}
Since the HiCOPS parallel processes run in SPMD fashion, the cost analysis for any parallel process (with variable input size) is applicable for the entire system. Also, the runtime cost for a parallel process ($p_{i}\ \epsilon\ P$) to execute superstep ($j$) can be modeled by only its local input size (i.e. database and dataset sizes) and available resources (i.e. number of cores, memory bandwidth). The parallel processes may execute the algorithmic work in a data parallel, task parallel or a hybrid task and data parallel model. As an example, the execution runtime (cost) for a parallel process $p_i$ to execute superstep ($j$) which first generates $D$ model-spectra using algorithm $k_1$ and then sorts them using algorithm $k_2$ in data parallel fashion (using all $c_{p_{i}}$ cores) will be given as follows:
\begin{equation}\label{eq:example}
    T_{j,p_{i}} = k_{j1}(D) + k_{j2}(D) + \gamma_{p_{i}}
\end{equation}

Similarly, if the above steps $k_{z}$ are performed in a hybrid task and data parallel fashion, the number of cores allocated to each ($k_{jz}$) must also be considered. For instance, in the above example, if the two algorithmic steps are executed in sub-task parallel fashion with $c_{p_{i}}/2$ cores each, the execution time will be given as:
\begin{equation}\label{eq:example2}
    T_{j,p_{i}} = max(k_{j1}(D,c_{p_{i}}/2), k_{j2}(D,c_{p_{i}}/2)) + \gamma_{p_{i}}
\end{equation}

For analysis purposes, if the time complexity of the algorithms used for step $k_{jz}$ is known (say $O(.)$), we will convert it into a linear function $k_{jz}'$ with its input data size multiplied by its runtime complexity. This conversion will allow better quantification of serial and parallel runtime portions as seen in later sections. As an example, if it is known that the sorting algorithms used for $k_{j2}$ have time complexity: $O(N \log N) $, the equation~\ref{eq:example} can be modified to:
\begin{equation}\label{eq:example3}
    T_{j,p_{i}} = k_{j1}(D) + k_{j2}'(D \log D) + \gamma_{p_{i}}
\end{equation}

\textbf{Remarks:} The formulated model will be used to analyze the runtime cost for each superstep, quantify the serial, parallel and overhead costs in the overall design, and optimize the overheads.

%
%
\subsection*{Superstep 1: Partial Database Construction}\label{sec:ss1}
In this superstep, the HiCOPS parallel processes construct a partial database by executing the following three algorithmic steps in data parallel fashion (Fig.~\ref{fig:main}):

\begin{enumerate}
\item Generate the whole peptide database and extract a (load balanced) partition.
\item Generate the model-spectra data\footnote{Currently only the b- and y-ions are generated.} from the local peptide database partition.
\item Index the local peptide and model-spectra databases (fragment-ion index).
\end{enumerate}

The database entries are generated and partitioned through the LBE algorithm \cite{haseeb2019lbe} supplemented with a new distance metric called \textit{Mod Distance} ($\Delta m$). The proposed $\Delta m$ separates the pairs of database entries based on the edit locations if they have the same Edit Distance ($\Delta e$) (See Supplementary Text~3). The reason for supplementing LBE with the new distance metric is to construct identical (load balanced) database partitions \cite{haseeb2019lbe} at parallel HiCOPS processes. Supplementary Fig.~3 illustrates the generic LBE algorithm, which to the best of our knowledge, is the only existing technique for efficient model-spectra database partitioning. For a pair of peptide database entries $(x,y)$, assuming the sum of unedited letters from both sequence termini is ($a$), the Mod Distance ($\Delta m$) is given as:
$$
\Delta m(x,y)= 2 - \frac{a}{max(len(x), len(y))}
$$

\textbf{Cost Analysis:} The first step generates the entire database of size ($D$) and extracts a partition (of roughly the size $D/P=D_{p_{i}}$) in runtime: $k_{11}(D)$. The second step generates the model-spectra from the partitioned database using standard the algorithms \cite{eng1994approach} in runtime: $k_{12}(D_{p_{i}})$. The third step constructs a fragment-ion index similar to \cite{kong2017msfragger}, \cite{chi2015pfind}, \cite{haseeb2019efficient} in runtime: $O(N \log N)$. In our implementation, we employed the CFIR-Index \cite{haseeb2019efficient} indexing method due to its smaller memory footprint. This results in time $k_{13}'(D_{p_{i}} \log D_{p_{i}})$ for the indexing step. Collectively, the runtime for this superstep is given by Equation~\ref{eq:1}.
\begin{dmath} \label{eq:1}
    T_{1} = max_{p_{i}}(k_{11}(D) + k_{12}(D_{p_{i}}) + k_{13}'(D_{p_{i}} \log D_{p_{i}}) + \gamma_{p_{i}})
\end{dmath}

\textbf{Remarks:} Equation~\ref{eq:1} depicts that the serial part of execution time i.e. $k_{11}(D)$ limits the parallel efficiency of superstep 1. However, using simpler but faster database partitioning may result in imbalanced partial databases leading to severe performance deprecation.

%
%
\subsection*{Superstep 2: Experimental MS/MS Data Pre-processing}
In this superstep, the HiCOPS parallel processes pre-process a partition of experimental MS/MS spectra data by executing the following three algorithmic steps in data parallel fashion (Fig.~\ref{fig:main}):

\begin{enumerate}
    \item Read the dataset files, create a batch index and initialize internal structures.
    \item Pre-process (i.e. normalize, clear noise, reconstruct etc.) a partition of experimental MS/MS data.
    \item Write-back the pre-processed data.
\end{enumerate}

The experimental spectra are split into batches such that a reasonable parallel granularity is achieved when these batches are searched against the database. By default, the maximum batch size is set to 10,000 spectra and the minimum number of batches per dataset is set to $P$. The batch information is indexed using a queue and a pointer stack to allow quick access to the pre-processed experimental data in the superstep 3.

\textbf{Cost Analysis:} The first step for reads the entire dataset (size: $q\beta$) and creates a batch index in runtime: $k_{21}(q\beta)$. The second step may pre-process a partition of the dataset (of roughly the size: $q\beta /P = Q_{p_{i}}$) using a data pre-processing algorithm such as \cite{ding2009novel}, \cite{deng2019pclean}, \cite{liu2020full} etc. in runtime: $k_{22}(Q_{p_{i}})$. The third step may write the pre-processed data back to the file system in runtime: $k_{23}(Q_{p_{i}})$. Note that the second and third steps may altogether be skipped in subsequent runs or in case when the pre-processed spectra data are available. Collectively, the runtime for this superstep is given by Equation~\ref{eq:2}.
\begin{dmath}\label{eq:2}
    T_{2} = max_{p_{i}}(k_{21}(q\beta) + k_{22}(Q_{p_{i}}) + k_{23}(Q_{p_{i}}) + \gamma_{p_{i}}) 
\end{dmath}

\textbf{Remarks:} Equation~\ref{eq:2} depicts that the parallel efficiency of superstep 2 is highly limited by its dominant serial portion i.e. $k_{21}(q \beta)$. Moreover, this superstep is sensitive to the file system bandwidth since large volumes of data may need to be read from and written to the shared file system.

%
%
\subsection*{Superstep 3: Partial Database Peptide Search} \label{sec:dbsrch}
This is the most important superstep in HiCOPS workflow and is responsible for 80-90\% of the database peptide search algorithmic workload in real world experiments. In this superstep, the HiCOPS parallel processes search the pre-processed experimental spectra against their partial databases by executing the following three algorithmic steps in a hybrid task and data parallel fashion (Fig.~\ref{fig:main}):

\begin{enumerate}
    \item Load the pre-processed experimental MS/MS data batches into memory.
    \item Search the loaded spectra batches against the (local) partial database and produce intermediate results.
    \item Serialize and write the intermediate results to the shared file system assigning them unique tags.
\end{enumerate}

Three parallel sub-tasks are created, namely $R$, $I$ and $K$, to execute the algorithmic work in this superstep in a producer-consumer pipeline. The data flow between parallel sub-tasks is handled through queues to create a buffer between producers and consumers. The first sub-task ($R$) loads batches of the pre-processed experimental spectra data and puts them in queue ($q_f$) as depicted in Supplementary Algorithm~\ref{alg:r}. The sub-task $R$ may also perform \emph{minimal} computations on the experimental spectra before putting them in queue. e.g. select only the ($B$) most-intense peaks from the experimental spectra. The parallel cores assigned to sub-task $R$ are given by: $\mid r\mid$. The second sub-task ($I$) extracts batches from ($q_f$), performs the database peptide search against its local database partition and puts the produced intermediate results in queue ($q_k$) depicted in Supplementary Algorithm~\ref{alg:i}. The parallel cores assigned to sub-task $I$ are given by: $\mid i\mid $. The sub-task $I$ also recycles the memory buffers back to sub-task $R$ for reuse, using the queue ($q_r$). The last sub-task ($K$) serializes and writes the intermediate results to the shared file system (or shared memory if available) using $\mid k\mid $ cores. Fig.~\ref{fig:grabs}c illustrates the pipeline setup in this superstep.

\textbf{Cost Analysis:} The sub-task ($R$) reads the experimental data batches in runtime: $k_{30}(q\beta)$. The sub-task ($I$) iteratively filters the partial database using multiple criteria followed by formal spectral comparisons (or scoring). Most commonly, the database peptide search algorithms use two or three database filtration steps such as peptide precursor mass tolerance \cite{mcilwain2014crux}, \cite{li2019sw}, shared fragment-ions \cite{kong2017msfragger}, \cite{chi2015pfind} and sequence tags \cite{chi2018open} \cite{devabhaktuni2019taggraph}. In our implementation, we use the first two filtration methods that execute in runtime: $k_{31}(qD_{p_{i}}) + k_{32}(q\beta\alpha_{p_{i}})$ respectively. Here, the $\alpha_{p_{i}}$ represents the average filtered database size filtered from the first step. The formal experimental spectrum to model-spectra comparisons (spectral comparisons) are performed using scoring methods such as cross-correlation \cite{eng1994approach}, hyperscore \cite{craig2003method} etc. in runtime: $k_{33}(q\beta\sigma_{p_{i}}) + k_{34}(q\mu_{p_{i}})$. Here, the $\sigma_{p_{i}}$ and $\mu_{p_{i}}$ represent the average number of filtered shared-ions and model-spectra per experimental spectrum. Finally, the sub-task $K$ writes the partial results to the shared file system in runtime: $k_{35}(q)$.

\textbf{Overhead Costs:} Multiple runtime overheads stemming from load imbalance, producer-consumer speed mismatch, file system bandwidth congestion can affect the performance of this superstep. Therefore, it is important to capture them using an additional runtime cost: $V_{p_{i}}(q, D_{p_{i}}, P)$. The optimizations implemented to alleviate these overhead costs in superstep 3 include buffering, task scheduling, load balancing and data sampling (discussed in later sections). Collectively, the runtime for this superstep is given by Equation~\ref{eq:6}.

The runtime of sub-task $R$, i.e. $t_{p_{i}}(r,\mid r\mid )$, is given as:
\begin{equation}\label{eq:3}
    t_{p_{i}}(r,\mid r\mid ) = k_{30}(q\beta, \mid r\mid)
\end{equation}

The runtime of sub-task $I$, i.e. $t_{p_{i}}(i,\mid i\mid )$, is given as:
$$
    t_{p_{i}}(i,\mid i\mid ) = k_{31}(qD_{p_{i}},\mid i \mid ) + k_{32}(q\beta\alpha_{p_{i}},\mid i\mid ) + k_{33}(q\beta\sigma_{p_{i}}) + k_{34}(q\mu_{p_{i}},\mid i\mid )
$$

Or:
\begin{equation}\label{eq:4}
    \begin{split}
        t_{p_{i}}(i,\mid i\mid) = k_{31}'(q\log(D_{p_{i}}),\mid i \mid) + k_{32}'(q\beta\log(\alpha_{p_{i}}), \mid i\mid) + \\ k_{33}(q\beta\sigma_{p_{i}},\mid i \mid ) + k_{34}(q\mu_{p_{i}},\mid i\mid)
    \end{split}
\end{equation}

The runtime of sub-task $K$, i.e. $t_{p_{i}}(k, \mid k\mid)$, is given as:
\begin{equation}\label{eq:5}
    t_{p_{i}}(k,\mid k\mid ) =  k_{35}(q,\mid k\mid )
\end{equation}

Combining equations \ref{eq:3}, \ref{eq:4} and \ref{eq:5} we have:
\begin{equation}\label{eq:6}
    \begin{split}
        T_{3} = max_{p_{i}}(max(t_{p_{i}}(r,\mid r\mid ), t_{p_{i}}(i,\mid i\mid ), t_{p_{i}}(k,\mid k\mid )) + \\V_{p_{i}}(q, D_{p_{i}}, P) + \gamma_{p_{i}})
    \end{split}
\end{equation}

\textbf{Remarks:} Equations~\ref{eq:3}, \ref{eq:4}, \ref{eq:5} and \ref{eq:6} depict that the parallel runtime portion of this superstep grows quadratically superseding the (small) serial portions capable of near ideal parallel efficiency if the overheads are eliminated.


%
%
\subsection*{Superstep 4: Result Assembly}
In this superstep, the HiCOPS parallel processes assemble the intermediate results from the last superstep into complete results by executing the following algorithmic steps in a hybrid task and data parallel fashion (Fig.~\ref{fig:main}d):

\begin{enumerate}
    \item Read a set of intermediate result batches, assemble them into complete results, and send the assembled results to their parent processes.
    \item Receive complete results from other parallel processes and synchronize communication.
    \item Write the complete results to the file system.
\end{enumerate}

Two parallel sub-tasks are created to execute the algorithmic steps in this superstep. The first sub-task reads sets of intermediate results from the shared file system (or shared memory) (satisfying: $tag \% p_{i} = 0$; $p_{i}\ \epsilon$ MPI ranks), de-serializes them and assembles the complete results. The statistical significance scores are then computed and sent to their origin processes. For example, the process with MPI rank 4 will process the all intermediate result batches with tag 0x8\_$i$ where $i=0,1,..,P-1$. The assembly process is done through signal addition and shift operations illustrated in Fig.~\ref{fig:grabs}d. The expectation scores (e-Values ($ev$)) are computed using null hypothesis approach by first smoothing the assembled data through Savitzky-Golay filter and then applying significance test through either the Linear-Tail Fit \cite{fenyo2003method} or log-Weibull (Gumbel) Fit method illustrated in Fig.~\ref{fig:grabs}d. The computed e-Values along with additional information (16 bytes) are sent to the HiCOPS process that recorded the most significant database hit (origin). The computed results are not sent immediately but are accumulated in a map data structure and sent collectively when all batches are done. All available cores ($c_{p_{i}}$) are assigned to this sub-task. Supplementary Algorithm~\ref{alg:assemble} depicts the algorithmic work performed by this sub-task.

The second sub-task runs waits for $P-1$ packets of complete data from other HiCOPS processes. This task runs inside an extra (over-subscribed) thread in a concurrent fashion and only activates when incoming data is detected. Finally, once the two sub-tasks complete (join), the complete results are written to the file system in data parallel fashion using all available threads.

\textbf{Cost Analysis:} The first sub-task reads the intermediate results, performs regression and sends computed results to other processes in runtime: $k_{41}(Q_{p_{i}},c_{p_{i}}) + k_{42}(Q_{p_{i}},c_{p_{i}}) + k_{43}(P, 1)$ time. The second sub-task receives complete results from other tasks in runtime: $k_{44}(P, 1)$. Finally, the complete results are written in runtime: $k_{45}(Q_{p_{i}})$. Collectively, the runtime for this superstep is given by equation~\ref{eq:7aa}.
\begin{dmath}\label{eq:7aa}
    T_{4} = max_{p_{i}}(max(k_{41}(Q_{p_{i}},c_{p_{i}}) + k_{42}(Q_{p_{i}},c_{p_{i}}) + k_{43}(P,1), k_{44}(P,1)) + k_{45}(Q_{p_{i}}) + \gamma_{p_{i}})
\end{dmath}

To simplify equation~\ref{eq:7aa}, we can re-write it as a sum of computation costs plus the communication overheads ($k_{com}(P,1)$) as:
\begin{dmath}\label{eq:7b}
    T_{4} = max_{p_{i}}(k_{41}(Q_{p_{i}},c_{p_{i}}) + k_{42}(Q_{p_{i}},c_{p_{i}}) +  k_{com}(P,1) + k_{45}(Q_{p_{i}}) + \gamma_{p_{i}})
\end{dmath}

Assuming that the network latency is denoted as ($\omega$), bandwidth is denoted as ($\pi$) and $(16Q_{p_{i}})$ is the average data packet size in bytes, the inter-process communication overhead cost ($k_{com}(P,1)$) in seconds is estimated to be:
$$
    k_{com}(P,1) \approx 2(P-1)(\omega + 16Q_{p_{i}}/\pi)
$$

\textbf{Remarks:} As the communication per process are limited to only one data exchange between any pair of processes, the overall runtime given by equation~\ref{eq:7b} is highly scalable. The effective communication cost depends on the amount of overlap with computations and the network parameters at the time of experiment.
%
%
\subsection*{Performance Analysis}
To quantify the parallel performance, we decompose the total HiCOPS time $T_{H}$ (Eq.~\ref{eq:tot}) into three runtime components. i.e. parallel runtime ($T_{p}$), serial runtime ($T_{s}$) and overheads runtime ($T_o$) given as:
\begin{equation}
    T_{H} = \sum_{j=1}^{4}{max_{p_{i}}(T_{j,p_{i}})} = T_{o} + T_s + T_p
\end{equation}

Using equations~\ref{eq:tot}, \ref{eq:1}, \ref{eq:2}, \ref{eq:6}, and \ref{eq:7b}, we separate the three runtime components as:

\begin{equation}\label{eq:to}
    T_{o} = V_{p_{i}}(q,D_{p_{i}}, P) + \gamma_{p_{i}}
\end{equation}
\begin{equation}
    T_{s} = k_{11}(D) + k_{21}(q\beta) + k_{com}(P,1)
\end{equation}
and:
\begin{equation}
    \begin{split}\label{eq:tp}
        T_{p} = k_{12}(D_{p_{i}}) + k_{13}'(D_{p_{i}} \log D_{p_{i}}) + k_{22}(Q_{p_{i}}) + k_{23}(Q_{p_{i}}) + \\ max(t_{p_{i}}(t, \mid r \mid), t_{p_{i}}(i, \mid i \mid), t_{p_{i}}(k, \mid k \mid)) + k_{41}(Q_{p_{i}},c_{p_{i}}) + \\
        k_{42}(Q_{p_{i}},c_{p_{i}}) + k_{45}(Q_{p_{i}})
    \end{split}
\end{equation}

$T_{s}$ is the minimum serial time required for HiCOPS execution and cannot be further reduced. Therefore, we will focus on optimizing the remaining runtime: $T_{F} = T_{p} + T_{o}$. Using equations~\ref{eq:to} and \ref{eq:tp}, we have:
\begin{equation}\label{eq:tf}
    \begin{split}
        T_{F} = k_{12}(D_{p_{i}}) + k_{13}'(D_{p_{i}} \log D_{p_{i}}) + k_{22}(Q_{p_{i}}) + k_{23}(Q_{p_{i}}) + \\ max(t_{p_{i}}(t, \mid r \mid), t_{p_{i}}(i, \mid i \mid), t_{p_{i}}(k, \mid k \mid)) + 
        k_{41}(Q_{p_{i}},c_{p_{i}}) + \\
        k_{42}(Q_{p_{i}},c_{p_{i}}) + k_{45}(Q_{p_{i}}) + T_{o}
    \end{split}
\end{equation}

Since the HiCOPS parallel processes divide the database and experimental dataset roughly fairly in supersteps 1 and 2, the first four and the sixth term in $T_{p}$ are already almost optimized, so we can prune them from $T_F$:
\begin{equation}
    \begin{split}
        T_{F} = max(t_{p_{i}}(t, \mid r \mid), t_{p_{i}}(i, \mid i \mid), t_{p_{i}}(k, \mid k \mid)) + k_{41}(Q_{p_{i}},c_{p_{i}}) + \\k_{42}(Q_{p_{i}},c_{p_{i}}) +  + k_{45}(Q_{p_{i}}) + T_{o}
    \end{split}
\end{equation}

Recall that the superstep 4 runtime is optimized for maximum parallelism (and least inter-process communication) and that the superstep 3 performs the largest fraction of overall algorithmic workload. Thus, we can also remove the superstep 4 terms from $T_{F}$ to simplify analysis:
$$
    T_{F} = max(t_{p_{i}}(t, \mid r \mid), t_{p_{i}}(i, \mid i \mid), t_{p_{i}}(k, \mid k \mid)) + T_{o}
$$

Further, as that the superstep 3 is executed using the producer-consumer pipeline (Fig.~\ref{fig:grabs}c) where the sub-task $R$ must produce all data before it can be consumed by $I$ meaning its runtime must also be smaller than $t_{p_{i}}(i, \mid i \mid)$ and $t_{p_{i}}(k, \mid k \mid)$ allowing a safe removal from the above equation yielding:
$$
    T_{F} = max(t_{p_{i}}(i, \mid i \mid), t_{p_{i}}(k, \mid k \mid)) + T_{o}
$$

In above equation, we can rewrite the $max(.)$ term as the time to complete sub-task $I$: ($t_{p_{i}}(i, \mid i \mid)$) plus the extra time to complete sub-task $K$ (the last consumer): $t_{x}(k)$. Therefore, using equation~\ref{eq:5} we have:
\begin{equation}\label{eq:2l}
    \begin{split}
        T_{F} = k_{31}'(q\log(D_{p_{i}}),\mid i \mid) + k_{32}'(q\beta\log(\alpha_{p_{i}}), \mid i\mid) + \\ k_{33}(q\beta\sigma_{p_{i}},\mid i \mid ) + k_{34}(q\mu_{p_{i}},\mid i\mid) + t_{x}(k) + T_{o}
    \end{split}
\end{equation}

We can prune the first two terms in the equation~\ref{eq:2l} as well since their runtime contribution: $O(\log N)$ will be relatively very small. Finally, using equation~\ref{eq:to} in \ref{eq:2l}, we have:  
\begin{equation}\label{eq:last}
    \begin{split}
        T_{F} = k_{33}(q\beta\sigma_{p_{i}},\mid i \mid ) + k_{34}(q\mu_{p_{i}},\mid i\mid) + t_{x}(k) + V_{p_{i}}(q,D_{p_{i}}, P) + \gamma_{p_{i}}
    \end{split}
\end{equation}

\textbf{Remarks:} The equations~\ref{eq:tf}-\ref{eq:2l} and the simplifications made may be modified according to the changes in superstep design and/or the algorithms employed in either superstep.

%
%
\subsection*{Optimizations}
The overhead cost term: $V_{i}(q,P)$ represents the load imbalance (or synchronization), producer-consumer speed mismatch, and data read costs, while the term: $t_{x}(k)$ represents the data write cost. Note that these overheads may result in a large subset of processing cores to idle (wasted CPU cycles). Furthermore, note that the load imbalance cost encapsulates all other costs in itself. The following sections discuss the optimization techniques employed to alleviate these overhead costs.

\subsubsection*{Buffering}
Four queues, the forward queue ($q_f$), recycle queue ($q_r$) and result queues ($q_k$, $q_{k}'$) are initialized and routed between the producer-consumer sub-tasks in the superstep 3 (Fig.~\ref{fig:grabs}c) as: $R \rightarrow I$, $R \leftarrow I$, $I \rightarrow K$ and $I \leftarrow K$ respectively. The $q_r$ is initialized with (default: 20) empty buffers for the sub-task $R$ to fill the pre-processed experimental data batches and push in $q_f$. The sub-task $I$ removes a buffer from $q_f$, consumes it (searches it) and pushes back to $q_r$ for re-use. The results are pushed to $q_k$ which are consumed by sub-task $K$ and pushed back to $q_{k}'$ for re-use. Three regions are defined for the queue $q_f$ based on the number of data buffers it contains at any time. i.e. $w_1:$ ($q_{f}.len$ $<$ 5), $w_2:$ (5 $\leq$ $q_{f}.len$ $<$ 15) and $w_3:$ ($q_f.len$ $\geq$ 15). These regions ($w_l$) are used by the task-scheduling algorithm discussed in the following section.

\subsubsection*{Task Scheduling} 
The task scheduling algorithm is used to maintain a synergy between the producer-consumer (sub-task) pipeline in the superstep 3. The algorithm initializes a thread pool of $c_{p_{i}} + 2$ threads where $c_{p_{i}}$ is the number of available cores. In the first iteration, 2 threads are assigned to the sub-tasks $R$ and $K$ while the remaining $c_{p_{i}} - 2$ threads are assigned to sub-task $I$. Then, in each iteration, the $q_f$ region: $w_{l}$, and the $q_f.pop()$ time for $I$, given by: $t_{wait}$, are monitored. A time series is built to forecast the next $t_{wait}$ (i.e. $t_{fct}$) using double exponential smoothing \cite{laviola2003double}. The $t_{wait}$ is also accumulated into $t_{cum}$. Two thresholds are defined: minimum wait ($t_{min}$) and maximum cumulative wait ($t_{max}$). Using all this information, a thread is removed from sub-task $I$ and added to $R$ if the following conditions are satisfied:
$$
    c_{I\rightarrow R} = (t_{wait} \geq t_{min} \wedge (t_{cum} + t_{fct}) > t_{max}) \vee (w_l = w_1\ \wedge \mid r\mid = 0)
$$

The $t_{cum}$ is set to 0 every time a thread is added to $R$. Similarly, a thread is removed from sub-task $R$ and added to $I$ if the following conditions are satisfied. All threads are removed from $R$ if the queue $q_f$ becomes full or there is no more experimental MS/MS data left to be loaded.
$$
    c_{R\rightarrow I} = (w_{l} = w_{3}\ \wedge \mid r\mid > 1) \vee q_{f}.full()
$$

The sub-task $K$ uses its 2 over-subscribed threads to perform the overlapped I/O operations concurrently (Fig.~\ref{fig:grabs}c).

\subsubsection*{Load Balancing} 
The algorithmic workload in equation~\ref{eq:last} is given by: $k_{33}(q\beta\sigma_{p_{i}},\mid i\mid) + k_{34}(q\mu_{p_{i}},\mid i\mid)$. Here, the terms $q\beta$ and $q$ are constants (experimental data size) whereas the terms $\sigma_{p_{i}}$ and $\mu_{p_{i}}$ are variable. The variable terms represent the filtered database size for a parallel HiCOPS process ($p_{i}$) and thus, must be balanced across processes. We do this statically by constructing balanced database partitions (hence a balanced workload) using the LBE algorithm supplemented with our new \textit{Mod Distance} metric in Superstep 1 (Online Methods, Fig.~\ref{fig:grabs}a, Supplementary Fig.~3). The correctness of the LBE algorithm for load balancing is proven in Supplementary Text~4. In future, we plan to devise and develop dynamic load balancing techniques in addition to this static technique for better results.

\subsubsection*{Sampling} 
\begin{figure}[ht]
    \centering
    \includegraphics[width=0.95\textwidth]{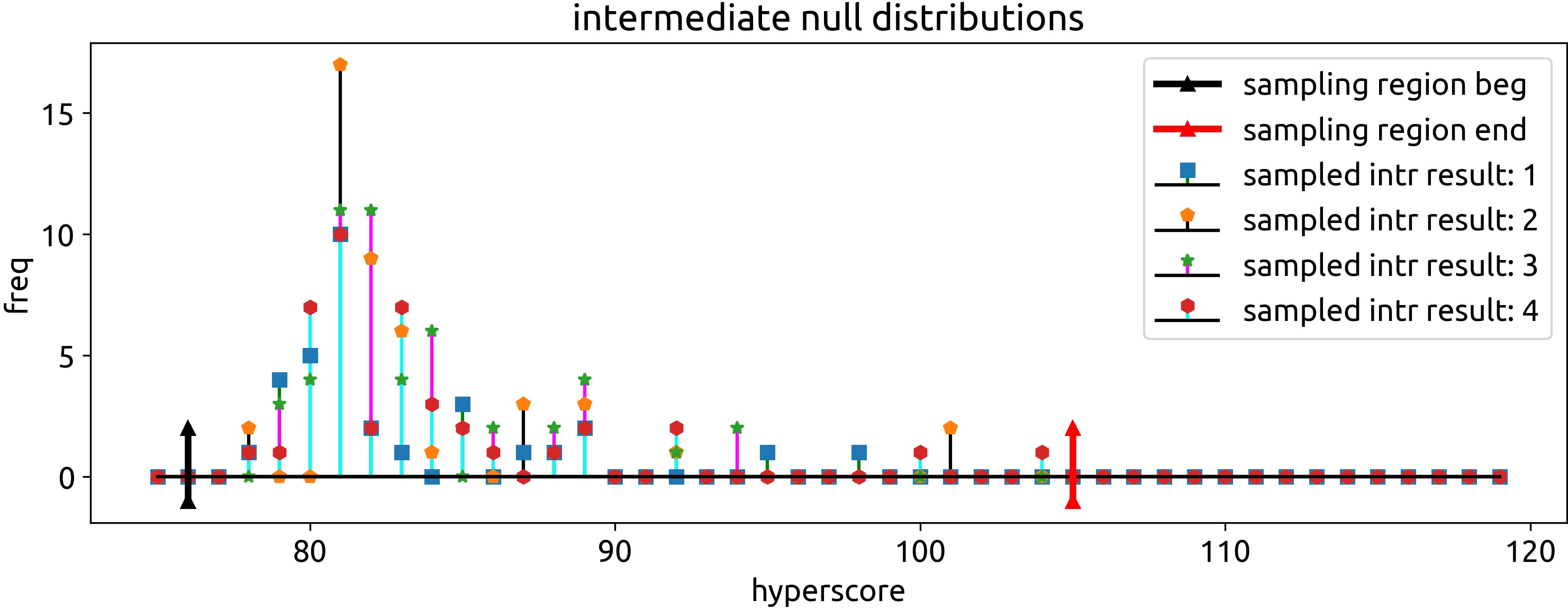}
    \caption{The intermediate results at all parallel processes are sampled around the mean. The mean is computed roughly by averaging the locations of three most intense samples in the distribution. Then, the most intense $s=$120 data points around the mean are kept around the mean and the others are discarded. The discarding method prunes the distribution tail samples first as they can be recovered by fitting a log-Weibull distribution in the sampled data.}
    \label{fig:sampling}
\end{figure}

The intermediate result produced by a parallel process ($p_{i}$) for an experimental spectrum ($q$) consists: $M$ top scoring database hits (8 bytes) and the frequency distribution of scores (local null distribution) (2048 bytes). Since this frequency distribution follows a log-Weibull, most of the data are localized near the mean. Using this information, we locate the mean and sample $s$ (default: 120) most intense samples from the distribution, and remove the samples, if necessary, from the tail first. This allows us to fit all the intermediate results in a buffer of 256 bytes limiting the size of each batch to 1.5MB. Thus, the intermediate results are almost instantly written to the file system by the sub-task $K$ resulting in minimum data write cost: $t_{x}(k)$. Fig.~\ref{fig:sampling} illustrates an example of the sampling method.

%
%
\subsection*{Code Availability}
The HiCOPS core parallel model and algorithms have been implemented using object-oriented C++14 and MPI. The rich instrumentation feature has been implemented via Timemory \cite{madsen2020timemory} for performance analysis and optimizations. Command-line tools for MPI task mapping (Supplementary Text~5, Supplementary Algorithm~\ref{alg:task}), user parameter parsing, peptide sequence database processing, file format conversion and result post-processing are also distributed with the HiCOPS framework. The build is managed via CMake 3.11+ \cite{hoffman2009software}. Please refer to the software web page: \textcolor{blue}{\url{hicops.github.io}} for source code and documentation.

%
%
\subsection*{Data Availability}
The datasets and database used in this study are publicly available from the mentioned respective data repositories. The experiment configuration files and raw results pertinent to the findings of this study are available from the corresponding author upon request.

%
%
\subsection*{Acknowledgments}
This work used the NSF Extreme Science and Engineering Discovery Environment (XSEDE) Supercomputers through allocations: TG-CCR150017 and TG-ASC200004. This research was supported by the NIGMS of the National Institutes of Health (NIH) under award number: R01GM134384. The authors were further supported by the National Science Foundations (NSF) under the award number: NSF CAREER OAC-1925960. The content is solely the responsibility of the authors and does not necessarily represent the official views of the National Institutes of Health and/or National Science Foundation.

\bibliographystyle{unsrt}

\bibliography{references}

\newpage
\setcounter{figure}{0}

\section*{Supplementary Figures}
\subsection*{Supplementary Figure 1}
The proteins are proteolyzed into peptides using an enzyme, typically Trypsin. The resultant peptide mixture is are fed to an automated liquid chromatography (LC) coupled two-staged MS/MS pipeline (LC-MS/MS) which yields the experimental MS/MS data.

\begin{figure}[ht]
    \centering
    \includegraphics[width=0.95\textwidth]{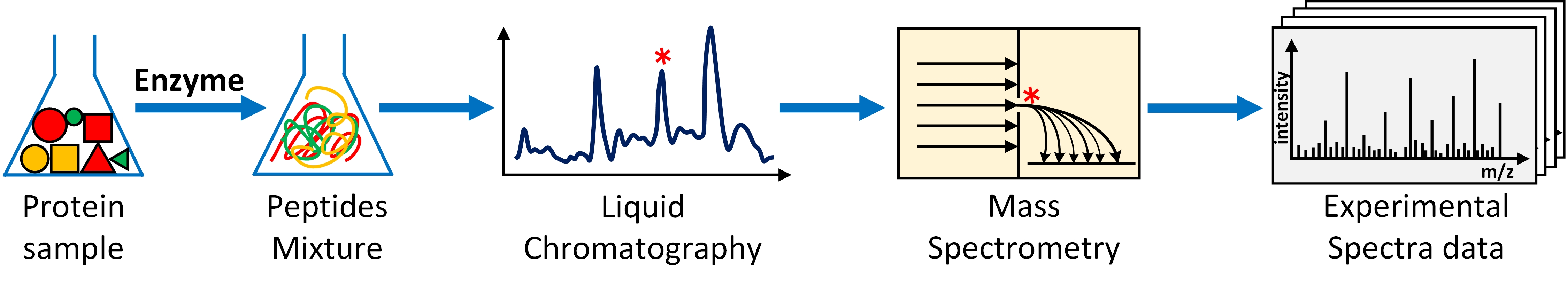}
    \label{fig:expt}
\end{figure}

\subsection*{Supplementary Figure 2}
The acquired experimental MS/MS data are compared against a database of model-spectra data. The model-spectra are simulated in-silico using a protein sequence database. Post-translational modifications (PTMs) are added in the simulation process to expand the search space.

\begin{figure}[ht]
    \centering
    \includegraphics[width=0.95\textwidth]{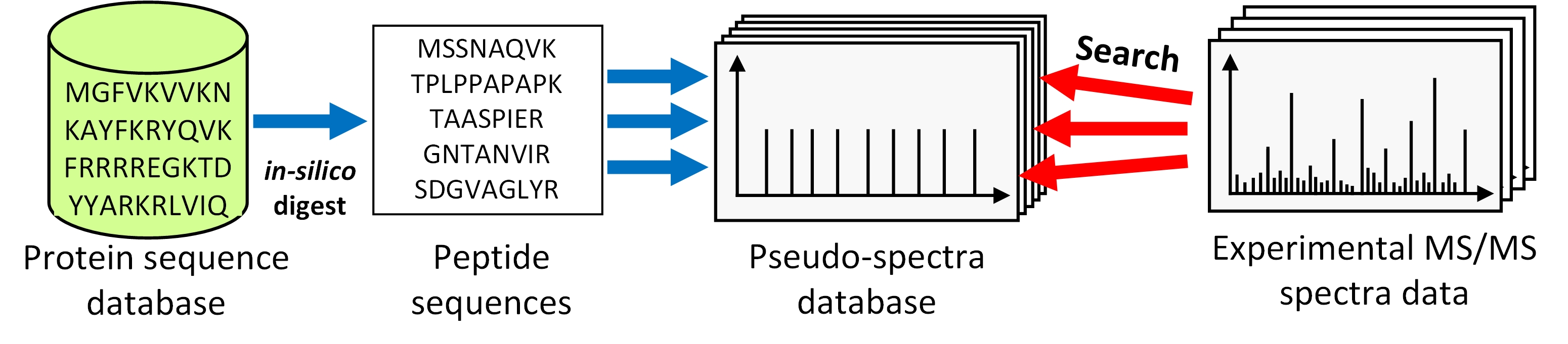}
    \label{fig:dbsearch}
\end{figure}

\subsection*{Supplementary Figure 3}
The improved LBE method used in the superstep 1 clusters the model-spectra database entries (shown as shapes) using two distance metrics: Edit Distance ($\Delta e$) and Mod Distance ($\Delta m$) (Supplementary Text~4). The obtained database clusters are then finely and evenly scattered across database partitions at parallel HiCOPS processes in either round robin or random fashion.

\begin{figure}[ht]
    \centering
    \includegraphics[width=0.90\textwidth]{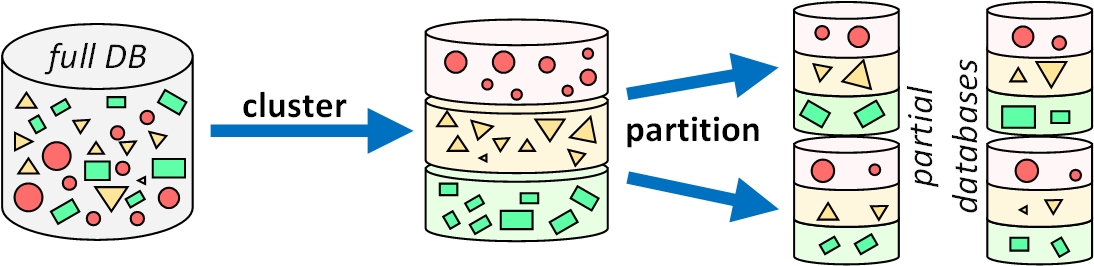}
\end{figure}

\newpage

\subsection*{Supplementary Figure 4}
The following sub-figures show the decomposition of the runtime, speedup and strong-scale efficiency results obtained for all 12 experiment sets ($e_{i}$) into individual supersteps ($s_{j}$) and overheads ($V$). The sub-figures depict that the overall efficiency increases as the workload (database, dataset and search filter) size increase. It can also be seen that the overall speedup (and efficiency) closely follows the superstep 3 ($s_{3}$) confirming its largest contribution towards the overall performance. This observation further indicates that the overheads associated with this supersteps must be correctly identified and optimized for the best performance. The observed super-linear speedups observed in case of large experimental workloads result from the improved CPU utilization due to the reduced memory intensity per parallel node (See Fig.~\ref{fig:speedups}c).

\begin{figure}[htpb]
    \centering
    \includegraphics[width=0.92\textwidth]{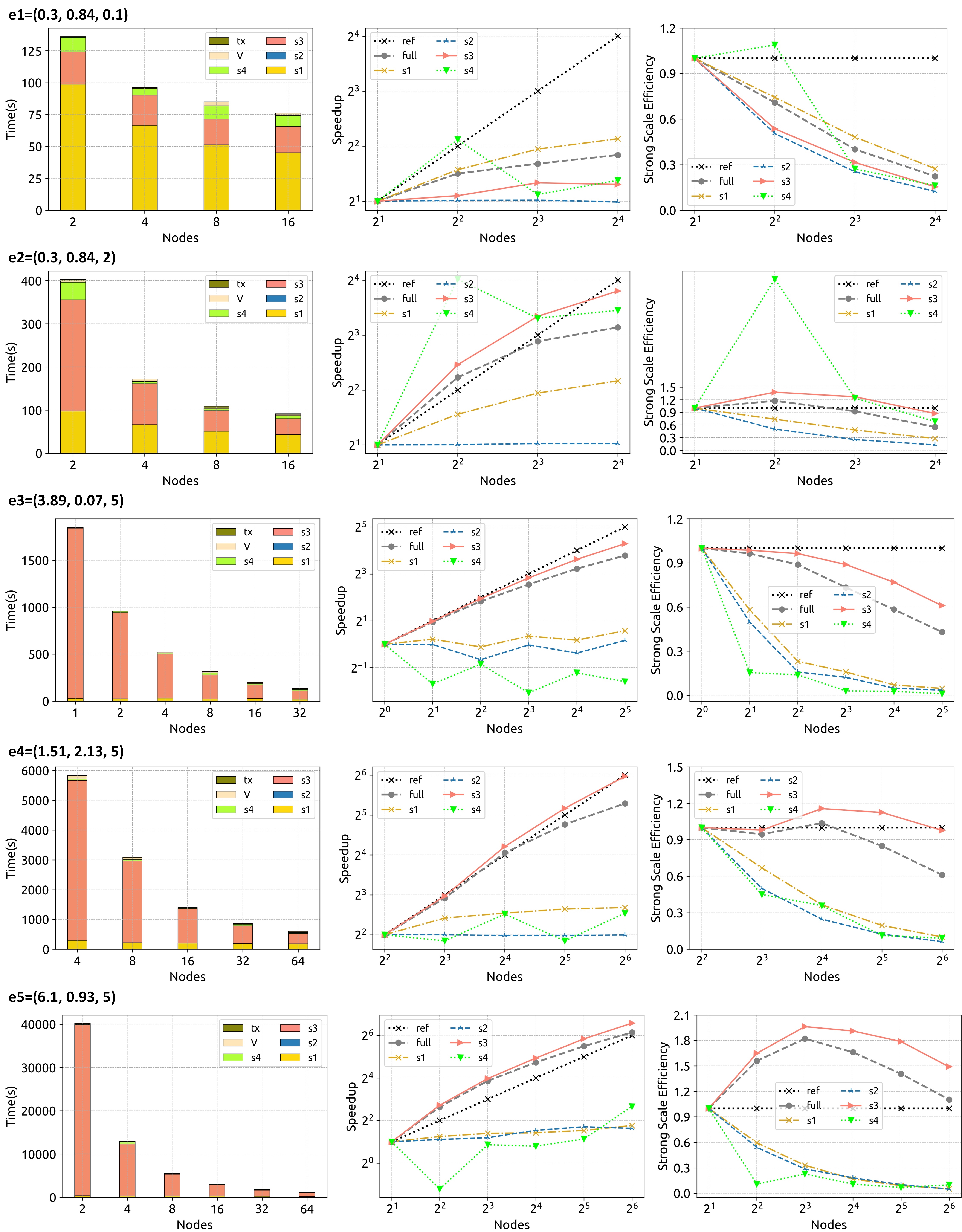}
    \label{fig:scale1}
\end{figure}

\begin{figure}[htpb]
    \centering
    \includegraphics[width=0.92\textwidth]{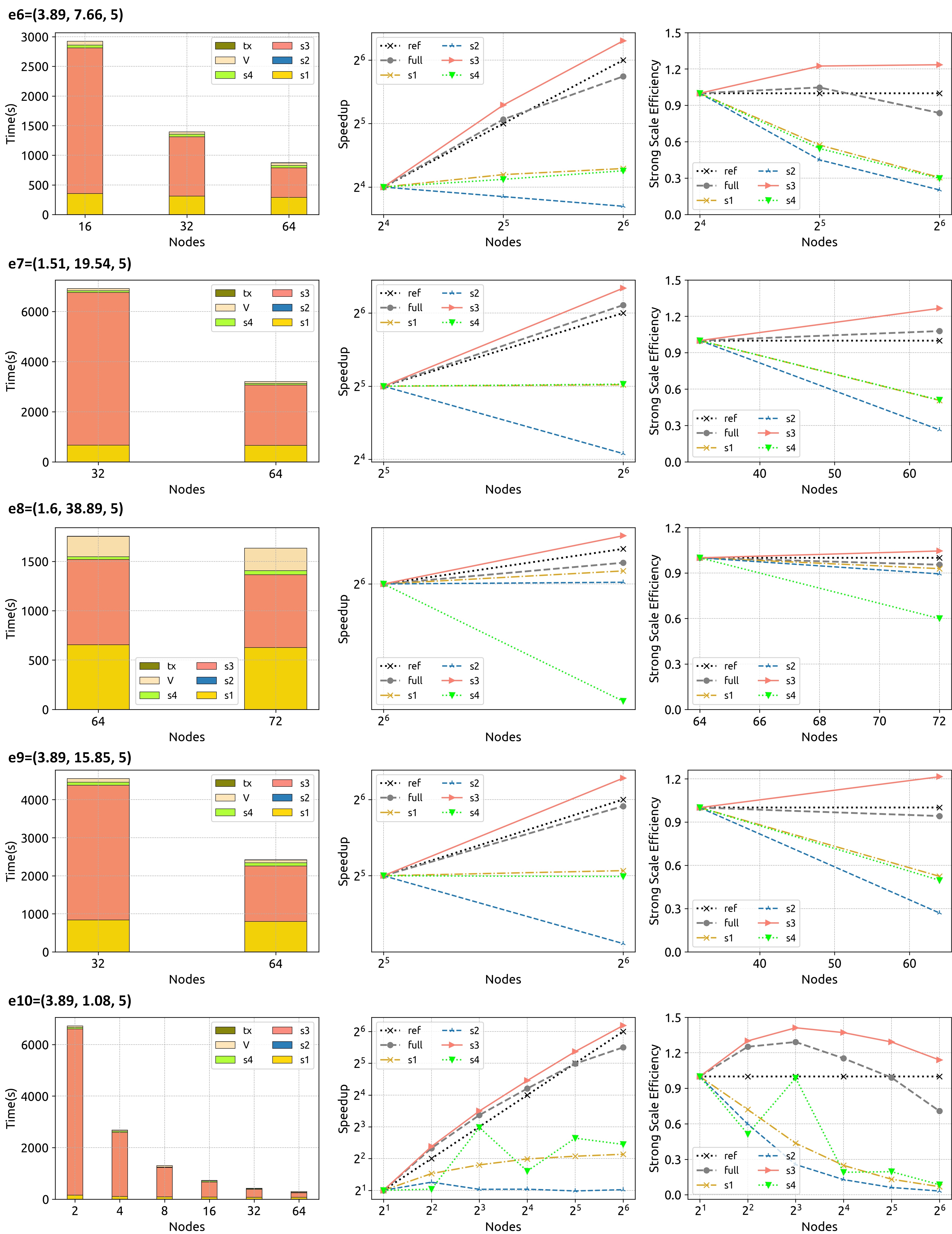}
    \label{fig:scale2}
\end{figure}

\begin{figure}[htpb]
    \centering
    \includegraphics[width=0.92\textwidth]{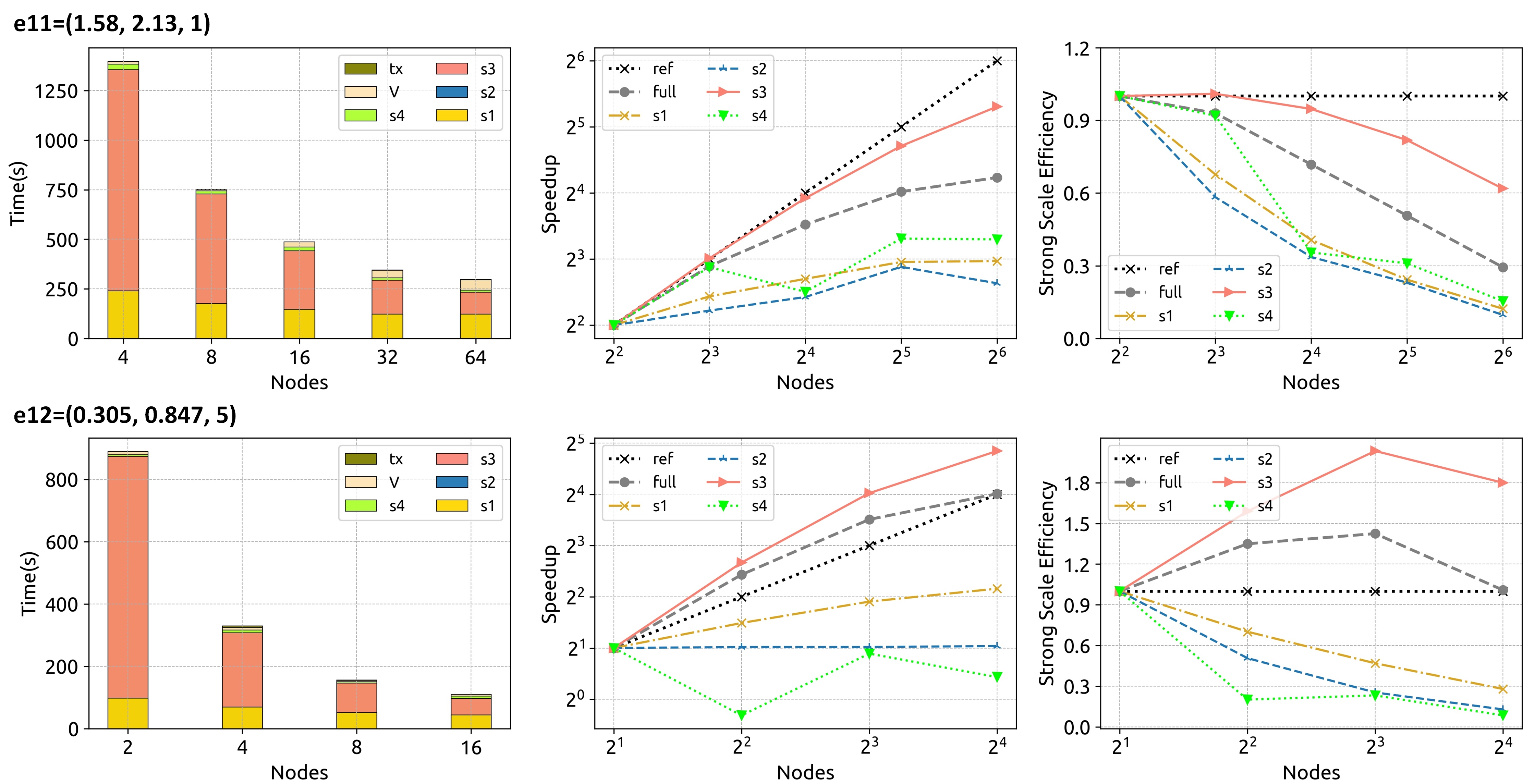}
    \label{fig:scale3}
\end{figure}

\newpage

\section*{Supplementary Text and Algorithms}

\subsection*{Supplementary Text 1}
\textbf{Related Work.} The distributed memory parallel database peptide search algorithms emerged with the Parallel Tandem \cite{duncan2005parallel}, which is a variant of the X!Tandem \cite{craig2004tandem} tool. Parallel Tandem achieves parallelism by spawning multiple instances of the original X!Tandem using MPI or PVM, where each instance processes a chunk of experimental dataset files. X!!Tandem \cite{bjornson2007x} is another variant of the X!Tandem which implements an internal (but similar) parallel technique for computational and synchronization steps. The experimental dataset files in the case of X!!Tandem are shuffled among the MPI processes to achieve better load balance. MR-Tandem \cite{pratt2011mr} follows a strategy similar to X!!Tandem however by breaking computations into small Map and Reduce tasks (Map-Reduce model) exhibiting better parallel efficiency than the Parallel Tandem and X!!Tandem. MCtandem \cite{li2019mctandem} and SW-Tandem \cite{li2019sw} implement the same parallel design but offload the X!Tandem's expensive Spectral Dot Product (SDP) computations over Intel Many Integrated Core (MIC) co-processor and Haswell AVX2 vector instructions respectively. Both algorithms also implement optimization techniques including double buffering, pre-fetching, overlapped communication and computations and a task-distributor for better performance. Bolt \cite{prakash2019bolt} implements a similar parallel design for MSFragger-like \cite{kong2017msfragger} algorithm where each parallel instance constructs a full model-spectra database and processes a chunk of experimental data.

\subsection*{Supplementary Text 2}
\textbf{Limitations in Related Work.} The major limitation in all existing distributed memory database peptide search algorithms is the inflated space complexity $=O(PN)$ where $P$ is the number of parallel nodes and $O(N)$ is the space complexity of their shared-memory counter parts. The space complexity inflation stems from the replication of massive model-spectra databases at all parallel instances. Consequently, the application of existing algorithms is limited to the use cases where the indexed model-spectra database size must fit within the main memory on all system nodes to avoid the expensive memory swaps, page faults, load imbalance and out-of-core processing overheads leading to an extremely inefficient solution. Furthermore, as the PTMs are added, this memory upper-bound is quickly exhausted due to the combinatorial increase in the database size \cite{chi2018open}, \cite{kong2017msfragger} incurring further slowdowns. For a reference, the model-spectra database constructed from a standard Homo sapiens proteome sequence database can grow from 3.8 million to 500 million model-spectra (0.6TB) if only the six most common PTMs (i.e. oxidation, phosphorylation, deamidation, acetylation, methylation and hydroxylation) are incorporated. There is some efforts towards  investigations of parallel strategies that involve splitting of model-spectra databases among parallel processing units \cite{kulkarni2009scalable}. In these designs, the database search is implemented in a stream fashion where each parallel process receives a batch of experimental data, executes partial search, and passes on the results to the next process in the stream. However, these models suffer from significant amounts of on-the-fly computations and frequent data communication between parallel nodes leading to high compute times, and limited ($\sim$50\%) parallel efficiency \cite{kulkarni2009scalable}.

\subsection*{Supplementary Text 3}
\textbf{Mod Distance.} The proposed \textit{Mod Distance} ($\Delta m$) is used as a supplementary metric in peptide database clustering in superstep 1 (the improved LBE method). The application of this metric can be best understood through an example. Consider three database peptide sequences $p$: \textcolor{blue}{MEGSYIRK}, $q$: \textcolor{blue}{M}\textcolor{red}{E*}\textcolor{blue}{GSY}\textcolor{red}{I*}\textcolor{blue}{RK} and $r$: \textcolor{blue}{MEG}\textcolor{red}{S*Y*}\textcolor{blue}{IRK}. The blue letters represent the normal amino acids in the peptide and the red letters with (*) represent the modified amino acids. Now, we can see that the Edit Distance between the pairs $\Delta e(p,q)=\Delta e(p,r)=2$ (cannot differentiate). Now let us apply the \textit{Mod Distance} on this scenario which considers the shared peaks between the peptide pairs to further separate them. For example, the shared (b- and y-) ions (or peaks) between $p$ and $q$ are: \textcolor{green}{M}\textcolor{red}{E*GSYI*}\textcolor{green}{RK} = 3 (green), yielding $\Delta m(p,q) = 1.625$ and the peaks shared between $p$ and $r$ are: \textcolor{green}{MEG}\textcolor{red}{S*Y*}\textcolor{green}{IRK} = 6 (green), yielding $\Delta m(p,r) = 1.25$. This indicates that the entries $p$ and $r$ should be located at relatively nearby database indices. The \textit{Mod Distance} can be easily generalized for other ion-series such as: a-, c-, x-, z-ions and immonium ions as well.

\subsection*{Supplementary Text 4}
\newtheorem*{theorem}{Correctness of LBE}

\begin{theorem}
Let the peptide precursor m/z distribution of any given database is $g(m)$ and that of any given dataset is $f(m)$, then the LBE algorithm statically results in fairly balanced workloads at all parallel nodes.

\begin{proof}
The algorithmic workload $w(f,g)$ for database peptide search can be given as the cost of performing the total number of comparisons to search the dataset $f(m)$ against the database $g(m)$ using filter size $\delta M$ and shared peaks $\geq k$, mathematically:
$$
w(f,g)= cost(\sum_{m=0}^{\infty}f(m) \sum_{z = -\delta M}^{\delta M} shp(f(m),g(m+z), k))
$$

where:
$$
shp(f,g,k) = count(shared\_peaks(f,g) \geq k)
$$

The above equations imply that the database distribution i.e. $\sum shp(f(m),g(m+z), k)$ must be similar at all parallel nodes in order to achieve system-wide load balance. The LBE algorithm achieves this by localizing (by $\delta M$ and shared peaks) the database entries and then finely scattering them across parallel nodes (Supplementary Fig.~3) producing identical local database distributions $g_{loc}(m)$ at parallel nodes thereby, identical workloads. This theorem can also be extended to incorporate \textit{sequence-tag} based filtration methods in a straightforward manner.
\end{proof}
\end{theorem}

\subsection*{Supplementary Text 5}
\textbf{Task Mapping.} The parallel HiCOPS tasks are configured and deployed on system nodes based on the available resources, user parameters and the database size. The presented algorithm assumes a Linux based homogeneous multicore nodes cluster where the interconnected nodes have multicores, local shared memory and optionally a local storage as well. This is the most common architecture in modern supercomputers including XSEDE Comet, NERSC Cori etc. The resource information is read using Linux's \texttt{lscpu} utility. Specifically, the information about shared memory per node ($\lambda$), NUMA nodes per node ($u$), cores per NUMA node ($c_{u}$), number of sockets per node ($s$) and cores per socket ($c_{s}$) is read. The total size of database ($D$) is then estimated using protein sequence database and user parameters. Assuming the total number of system nodes to be $P$, the parameters: number of MPI tasks per node ($t_{n}$) and the number of parallel cores per MPI task ($t_{c}$) and MPI task binding level ($t_{bl}$) are optimized as depicted in Supplementary Algorithm~\ref{alg:task}. The optimizations ensure that: 1) System resources are efficiently utilized 2) The MPI tasks have sufficient resources to process the database and 3) The MPI tasks have an exclusive access to a disjoint partition of local compute and memory resources.

Note that in Supplementary Algorithm~\ref{alg:task}, the lines 8 to 14 iteratively reduce the cores per MPI task while increasing the number of MPI tasks until the database size per MPI task is less than 48 million (empirically set for XSEDE Comet nodes). This was done to reduce the memory contention per MPI process for superior performance. The while loop may be removed or modified depending on the database search algorithms and machine parameters.

\subsection*{Supplementary Text 6}
\textbf{SW-Tandem.} The SW-Tandem binaries were obtained from its GitHub repository: \textcolor{blue}{\url{https://github.com/Logic09/SW-Tandem}} and were run on XSEDE Comet system with increasing number of parallel nodes using MPI but no speedups in runtime were observed. We repeatedly tried to contact the corresponding author about this issue via Email and GitHub issues (\textcolor{blue}{\url{https://github.com/Logic09/SW-Tandem/issues}}) but did not receive a response as of the submission date of this paper.

\newpage

\subsection*{Supplementary Algorithm 1}
\begin{algorithm}[thpb]
	\caption{Partial Database Construction in Superstep 1}
	\label{alg:ss1}
	\KwData{peptide sequences ($\epsilon$)}
	\KwResult{indexed partial database ($D_i$)}

    \tcc{generate database entries}
    \ForPar{$s \ in \ \epsilon$}
    {
        \For {$v \ in \ 2^{m} $}
        {
            $e \leftarrow gen\_entry(v)$\;

            \tcc{add to partial database if mine}
            \If{$is\_mine(e_{v})$}
            {
                $E.append(e)$\;
            }
        }
    }

    \tcc{generate model-spectra}
    \ForPar{$s \ in \ D_{i}$}
    {
        $S.append(model\_spectrum(s))$\;
    }

    \tcc{index the database in parallel}
    $D_{i} \leftarrow map(parallel\_sort(E), parallel\_index(S))$\;
    
    \tcc{return the indexed parital database}
    \Return $D_{i}$\;
\end{algorithm}

\newpage

\subsection*{Supplementary Algorithm 2}
\begin{algorithm}[hpbt]
	\caption{Data load (per thread) by sub-task $R$ (Superstep 3)}
	\label{alg:r}
	\KwData{forward queue ($q_f$), recycle queue ($q_r$), pointer stack ($s_d$), batch index ($i_d$)}

    \tcc{loop unless $q_{f}$ full, preempted or no more batches}
    \While{$\sim q_f.full()$}
    {
        \tcc{check pointer stack}
        \If {$\sim dp$}
        {
        $dp \leftarrow s_d.pop()$\;
        }
        
        \tcc{if stack is empty, get a new pointer}
        \If {$\sim dp$}
        {
            $dp \leftarrow i_d.pop()$\;
        }

        \tcc{no more experimental data batches - exit}
        \If {$\sim dp$}
        {
            $break$\;
        }
        
        \tcc{check preemption state and $q_r$ status}
        \eIf {$\sim preempt() \ or\ \sim q_r.empty()$}
        {
            $s_d.push(dp)$\;
            $break$\;
        }
        {
            \tcc{else get a buffer from $q_r$}
            $bp \leftarrow q_r.pop()$\;
        }
        
        \tcc{read a batch of expt data}
        $dp.read\_batch(bp)$\;
        
        \tcc{push the buffer to $q_f$}
        $q_f.push(bp)$\;
    }
\end{algorithm}

\newpage

\subsection*{Supplementary Algorithm 3}

\begin{algorithm}[hpbt]
	\caption{Partial DB search by sub-task $R$ (Superstep 3)}
	\label{alg:i}
	\KwData{forward queue ($q_f$), recycle queue ($q_r$), partial database ($D_{p_{i}}$), result queue ($q_k$)}

    \tcc {extract a batch from queue}
    $b \leftarrow q_f.pop()$\;
    
    \tcc{data parallel search}
    \ForPar{$e\ in\ b$}
    {
        \tcc{apply the precursor mass filter}
        $\sigma_{p_{i}} \leftarrow filter_{1}(D_{p_{i}}, e)$\;

        \If{$\sigma_{p_{i}}$}
        {
            \For{$\beta\ in\ e$}
            {
                \tcc{apply the shared peaks filter}
                $\mu_{p_{i}}.append(filter_{2}(\sigma_{p_{i}}, \beta))$\;
            }
            
            \tcc{score against the filtered database}
            \For{$h\ in\ \mu_{p_{i}}$}
            {
                $heap.push(k \leftarrow score(h,e))$\;
            }

            \tcc{append to a batch of intermediate results}
            $res_{i}.append(heap)$\;
        }
    }
    \tcc{recycle the buffer back to $q_r$}
    $q_r.push(b)$\;

    \tcc{push the intermediate results batch to $q_k$}
    $q_k.push(res_{i})$\;

\end{algorithm}

\newpage

\subsection*{Supplementary Algorithm 4}

\begin{algorithm}[htpb]
	\caption{Result Assembly in Superstep 4}
	\label{alg:assemble}
	\KwData{rank $p_{i}$, Intermediate Result batches ($r_{i}$)}
    \KwResult{expect scores ($ev$)}
    \tcc {extract a batch from queue}
    $b \leftarrow q_f.pop()$\;
    
    \tcc{get batches that satisfy the condition}
    \For{$b\ in\ (b\mod p_{i} = 0)$}
    {
            $l.append(b)$\;
    }
    \tcc{data parallel assembling of results for each batch}
    \ForPar{$s\ in\ l$}
    {
        \tcc{assemble the null distribution}
        $dist \leftarrow assemble(s)$\;

        \tcc{max heapify the scores}
        $heap \leftarrow make\_heap(s)$\;
        
        \tcc{use either fitting method}
        $fit \leftarrow logWeibullFit(dist)$\;
        $fit \leftarrow TailFit(dist)$\;
        
        \tcc{get the top hit from heap}
        $g_{max} \leftarrow heap.pop().value()$\;

        \tcc{compute the expect score}
        $ev \leftarrow (fit.w \times g_{max} + fit.b) \times heap.size()$\;
        
        \tcc{push results to a map structure}
        $map.push(key=g_{max}.key(), val=ev)$\;
    }

    \tcc{asynchronous scatter complete result data}
    \ForPar{$p_{i}\ in\ P$}
    {
       $isend(map.data(key = p_{i}), dst = p_{i})$\;
    }
    
    \tcc{synchronize using barrier}
    $barrier()$\;

    \tcc{write the results to the file system}
    $write(map.data(key = rank))$\;

\end{algorithm}

\newpage

\subsection*{Supplementary Algorithm 5}
\label{app:map}

\begin{algorithm}[hpbt]
	\caption{Task Mapping}
	\label{alg:task}
	\KwData{number of nodes ($n$), node parameters ($\lambda, u, c_{u}, s, c_{s}$) and database size ($D$)}
	\KwResult{number of MPI tasks per node ($t_{n}$), cores per MPI task ($t_{c}$) and MPI binding level ($t_{bl}$)}

    \tcc{ensure enough memory for database}
	\If{$D_{p_{i}} \leftarrow D/P > 0.70\lambda$}{
		\Return $err$\;
	}

    \tcc{set MPI binding level}
    $t_{bl} \leftarrow max\{u,c\}$\;

    \tcc{set MPI binding policy}
    $t_{bp} \leftarrow scatter$\;

    \tcc{set cores per MPI task}
    $t_{c} \leftarrow min\{c_{u}, c_{s}\}$\;

    \tcc{set number of MPI tasks per node}
    $t_{n} \leftarrow max\{u,c\}$\;

    $t_{max} \leftarrow t_{c}$\;

    \tcc{Optional: optimize for memory bandwidth}
    \While {$(D/t_{n} > 48 \times 10^{6})$}
    {
        \tcc{Choose the next highest factor of $t_{max}$}
        $n_{poss} \leftarrow factorize(t_{max})$\;
        \eIf{$n_{poss} \geq t_{max}/2$}
        {
            $t_{n} \leftarrow t_{n}\times t_{max}/n_{poss}$\;
            $t_{c} \leftarrow n_{poss}$\;
        }
        {
            $break$\;
        }
    }
	\Return $t_{n}, t_{c}, t_{bl}, t_{bp}$\;
\end{algorithm}

\newpage

\section*{Supplementary Tables}

\subsection*{Supplementary Table 1}
A snippet of the peptide-to-spectrum matches (PSMs) and e-values obtained by searching the dataset: $E_1$ against database: $D_1$ (no mods, $\delta M$=500Da). Full table can be requested from the corresponding author.
\begin{table}[htpb!]
  \centering
  \label{table:correct}
  \begin{tabular}{|c|c|c|c|c|c|}
    \hline
    \multirow{2}{*}{\parbox{2cm}{\centering{\textbf{ Matched Peptide}}}} & \multicolumn{5}{c|}{\textbf{e-Values for Parallel nodes}}\\
    \cline{2-6}
     & \textbf{1} & \textbf{2} & \textbf{4}, \textbf{8} & \textbf{16}, \textbf{32} & \textbf{64}\\
    \hline
    HLTYENVER & 6.6e-5 & 6.5e-5 & 6.5e-5 & 6.5e-5 & 6.5e-5\\
    \hline
    SEGESSRSVR & 3.175e-3  & 3.174e-3 & 3.174e-3 & 3.175e-3 & 3.174e-3\\
    \hline
    IFQCNKHMK & 0.037038  & 0.037037 & 0.037037 & 0.037036 & 0.037037\\
    \hline
    FIVSKNK & 0.113302  & 0.113301 & 0.113298 & 0.113297 & 0.113297\\
    \hline
    QQIVSGR & 1.294027  & 1.293975 & 1.293975 & 1.293975 & 1.293975\\
    \hline
    STVASMMHR & 2.641636  & 2.64151 & 2.64151 & 2.64151 & 2.64151\\
    \hline
    TLFKSSLK & 7.000016 & 7.0 & 7.0 & 7.0 & 7.0\\
    \hline
    QKQLLKEQK & 16.856401 & \multicolumn{2}{c|}{invalid} & 16.855967 & invalid\\
    \hline
\end{tabular}
\end{table}

\subsection*{Supplementary Table 2}
Speed comparison between existing tools and HiCOPS for the experiment 1a, dataset size: 8K, database size: 93.5M, precursor mass tolerance: $\delta M=$ 10.0Da.
\begin{table}[htpb!]
  \centering
  \label{table:e2a}
  \begin{tabular}{|c|c|c|c|c|c|}
    \hline
    \multirow{2}{*}{\parbox{2cm}{\centering{\textbf{Search Tool}}}} & \multicolumn{5}{c|}{\textbf{Execution Time (s) for parallel nodes}}\\
    \cline{2-6}
     & \textbf{1} & \textbf{2} & \textbf{4} & \textbf{8} & \textbf{16}\\
    \hline
    HiCOPS & - & 166.32 & 126.35 & 113.53 & 134.86\\
    \hline
    X!!Tandem & 4980 & 2445 & 1279.8 & 690 & 360\\
    \hline
    SW-Tandem & 1015 & 992 & 1002 & 999 & 1019\\
    \hline
    MSFragger & 299.4 & \multicolumn{4}{c|}{-}\\
    \hline
    X!Tandem & 957 & \multicolumn{4}{c|}{-}\\
    \hline
    Crux/Tide & 2470 & \multicolumn{4}{c|}{-}\\
    \hline
\end{tabular}
\end{table}

\newpage

\subsection*{Supplementary Table 3}
Speed comparison between existing tools and HiCOPS for the experiment 1b, dataset size: 8K, database size: 93.5M, precursor mass tolerance: $\delta M=$ 500.0Da\footnote{Tide limits the max peptide precursor tolerance to $\delta M$=$\pm$100Da}.
\begin{table}[htpb!]
  \centering
  \label{table:e2b}
  \begin{tabular}{|c|c|c|c|c|c|c|c|}
    \hline
    \multirow{2}{*}{\parbox{2cm}{\centering{\textbf{Search Tool}}}} & \multicolumn{7}{c|}{\textbf{Execution Time (s) for parallel nodes}}\\
    \cline{2-8}
     & \textbf{1} & \textbf{2} & \textbf{4} & \textbf{8} & \textbf{16}& \textbf{32}& \textbf{64}\\
    \hline
    HiCOPS & - & 188 & 135 & 115 & 101 & 101 & 144\\
    \hline
    X!!Tandem & 115K & 57.7K & 29.05K & 14.6K & 7.4K & 3.72K & 1.98K\\
    \hline
    SW-Tandem & 19.99K & 17.1K & 15.4K & 14.3K & 15.1K & 15K & 15K\\
    \hline
    MSFragger & 521 & \multicolumn{6}{c|}{-}\\
    \hline
    X!Tandem & 18.65K & \multicolumn{6}{c|}{-}\\
    \hline
    Crux/Tide & \multicolumn{7}{c|}{segmentation fault}\\
    \hline
\end{tabular}
\end{table}

\subsection*{Supplementary Table 4}
Speed comparison between HiCOPS and existing tools for the experiment 2a, dataset size: 3.8M, database size: 93.5M, precursor mass tolerance: $\delta M=$ 10.0Da. X!!Tandem and SW-Tandem ran for 2 days in all parallel configurations but failed to complete and were terminated by SLURM due to max job time limit on XSEDE Comet system.
\begin{table}[htpb!]
  \centering
  \label{table:e1a}
  \begin{tabular}{|c|c|c|c|c|c|}
    \hline
    \multirow{2}{*}{\parbox{2cm}{\centering{\textbf{Search Tool}}}} & \multicolumn{5}{c|}{\textbf{Execution Time (s) for parallel nodes}}\\
    \cline{2-6}
     & \textbf{1} & \textbf{2} & \textbf{4} & \textbf{8} & \textbf{16}\\
    \hline
    HiCOPS & - & 557.549 & 371.585 & 262.16 & 213.622 \\
    \hline
    X!!Tandem & \multicolumn{5}{c|}{terminated after 2 days}\\
    \hline
    SW-Tandem & \multicolumn{5}{c|}{terminated after 2 days}\\
    \hline
    MSFragger & 13402.66 & \multicolumn{4}{c|}{-} \\
    \hline
    X!Tandem & 1.71M & \multicolumn{4}{c|}{-}\\
    \hline
    Crux/Tide & 875.5K & \multicolumn{4}{c|}{-}\\
    \hline
\end{tabular}
\end{table}

\newpage 

\subsection*{Supplementary Table 5}
Speed comparison between HiCOPS and existing tools for the experiment 2b, dataset size: 3.8M, database size: 93.5M, precursor mass tolerance: $\delta M=$ 500.0Da\footnote{Tide limits the max peptide precursor tolerance to $\delta M$=$\pm$100Da}. X!!Tandem and SW-Tandem ran for 2 days in all parallel configurations but failed to complete and were terminated by SLURM due to max job time limit on XSEDE Comet system. X!Tandem has been running for 75 days at the time of submission of this manuscript and is expected to run over 8 months to complete its execution.
\begin{table}[htpb!]
  \centering
  \label{table:e1b}
  \begin{tabular}{|c|c|c|c|c|c|c|c|}
    \hline
    \multirow{2}{*}{\parbox{2cm}{\centering{\textbf{Search Tool}}}} & \multicolumn{7}{c|}{\textbf{Execution Time (s) for parallel nodes}}\\
    \cline{2-8}
    & \textbf{1} & \textbf{2} & \textbf{4} & \textbf{8} & \textbf{16}& \textbf{32}& \textbf{64}\\
    \hline
    HiCOPS & - & 23.5K & 6.6K & 2.8K & 1.4K & 807 & 485\\
    \hline
    X!!Tandem & \multicolumn{7}{c|}{terminated after 2 days}\\
    \hline
    SW-Tandem & \multicolumn{7}{c|}{terminated after 2 days}\\
    \hline
    MSFragger & 170.1K & \multicolumn{6}{c|}{-}\\
    \hline
    X!Tandem & 75 days* & \multicolumn{6}{c|}{-}\\
    \hline
    Crux/Tide & \multicolumn{7}{c|}{segmentation fault}\\
    \hline
\end{tabular}
\end{table}

\newpage 

\section*{Supplementary Protocol}
\subsection*{Minimum Environment}
\begin{itemize}
    \item Laptop, desktop, SMP cluster (HPC) with Linux OS
    \item GCC 7.2+ compiler with C++14, OpenMP and threading
    \item MPI with multiple threads support
    \item Python 3.7+ and common packages
    \item CMake 3.11+
\end{itemize}

\subsection*{Install}
Comprehensive details about the required packages, supported environments and step by step installation of packages and HiCOPS are documented at: \textcolor{blue}{\url{hicops.github.io/installation}}. This link will be updated as the development progresses.

\subsection*{Getting Started}
The instructions for setting up the peptide database, experimental MS/MS dataset and running HiCOPS are documented at: \textcolor{blue}{\url{hicops.github.io/getting_started}}. If you are running HiCOPS on SDSC XSEDE Comet cluster, you can follow a simpler set of instructions documented at: \textcolor{blue}{\url{hicops.github.io/getting_started/xsede}}

\subsection*{Integrating with HiCOPS framework}
The details on integrating the existing and new algorithms with the HiCOPS parallel core library are documented at: \textcolor{blue}{\url{hicops.github.io/getting_started/integrate}}. Currently, the integration must be done via the provided functional interface (and data structures). In near future, the integration will be redesigned using C++ template meta-programming interface. The documentation will be updated accordingly.

\subsection*{Command-line tools}
Several command-line tools are distributed as a part of HiCOPS software. These tools provide support for runtime interface, preparation of database, dataset, and post-processing final results. A brief summary of each tool is documented at: \textcolor{blue}{\url{hicops.github.io/tools}}. More tools will be provided in the future releases.

\subsection*{Current version}
The current released version of HiCOPS is v1.0: \textcolor{blue}{\url{hicops.github.io}}

\end{document}